\documentclass[twocolumn, twocolappendix, numberedappendix]{openjournal}

\usepackage{float}
\usepackage{subcaption}
\usepackage{amsmath}

\usepackage{xcolor}
\usepackage{footmisc}
\usepackage{textgreek}
\usepackage[utf8]{inputenc}
\usepackage[english]{babel}
\usepackage{natbib}
\usepackage{caption}

\usepackage{hyperref}
\hypersetup{
    unicode,
    colorlinks=true,
    linkcolor=linkcolor,
    citecolor=linkcolor,
    filecolor=linkcolor,
    urlcolor=linkcolor,
}
\usepackage{color,colortbl}
\definecolor{linkcolor}{rgb}{0.0,0.3,0.5}
\usepackage{tensind}
\tensordelimiter{?}
\DeclareGraphicsExtensions{.bmp,.png,.jpg,.pdf}
\usepackage{verbatim}
\usepackage[normalem]{ulem}
\usepackage{orcidlink}
\usepackage{soul}
\usepackage{xcolor}

\newcommand{\LCDM}{$\Lambda$CDM}
\newcommand{\wowa}{$w_0w_a$CDM}
\newcommand{\nz}{$n(z)$}
\newcommand{\txtp}{3$\times$2pt}

\graphicspath{ {./figs/} }

\begin{document}
\title{On-the-Fly Gaussian Process Reconstruction of the Galaxy Redshift Distribution for 3$\times$2pt Cosmology}

\author{Jip de Buck\orcidlink{0009-0001-5175-9282}}
\email{buck@strw.leidenuniv.nl}
\affiliation{Leiden Observatory, Leiden University, Einsteinweg 55, 2333 CC Leiden, The Netherlands}

\author{Guadalupe Cañas-Herrera\orcidlink{0000-0003-2796-2149}}
\email{canasherrera@strw.leidenuniv.nl}
\affiliation{Leiden Observatory, Leiden University, Einsteinweg 55, 2333 CC Leiden, The Netherlands}

\author{Henk Hoekstra\orcidlink{0000-0002-0641-3231}}
\email{hoekstra@strw.leidenuniv.nl}
\affiliation{Leiden Observatory, Leiden University, Einsteinweg 55, 2333 CC Leiden, The Netherlands}

\author{Angus H. Wright\orcidlink{0000-0001-7363-7932}}
\email{awright@astro.rub.de}
\affiliation{Astronomical Institute (AIRUB), Faculty of Physics and Astronomy, Ruhr University Bochum, German Centre for Cosmological Lensing, 44780 Bochum, Germany}

\author{}

\begin{abstract}
The joint analysis of weak gravitational lensing, photometric galaxy clustering, and their
cross-correlation, the so-called \txtp{} analysis, is one of the primary probes Stage-IV
surveys will use to constrain the dark-energy equation of state. Its accuracy depends critically on the calibration of the underlying photometric redshift distributions,
\nz{}, whose residual uncertainties can bias cosmological inference. However, conventional parametric nuisance models may not adequately capture the full range of plausible \nz{} uncertainties. We therefore present a cosmological inference framework that instead reconstructs \nz{} on the fly
using Gaussian Processes (GPs), replacing per-bin shift-and-width parameterisations by a
non-parametric, data-driven model. We validate the method on
synthetic \textit{Euclid}-like \txtp{} data vectors representative of \textit{Euclid} Data Release 1, sampling GP hyperparameters jointly with
cosmological and nuisance parameters to reliably propagate \nz{} uncertainties into the posterior distributions. We find that the dependence of the inferred parameters on the adopted \nz{} model is
markedly stronger in \wowa{} than in \LCDM{}. Opening the dark-energy sector both amplifies the
biases induced by a given \nz{} treatment and changes which \nz{} reconstruction the data prefer,
so that a model validated in \LCDM{} does not automatically remain adequate once $w_0$ and $w_a$
are free. Our results thus indicate that Stage-IV surveys should define \nz{} uncertainty marginalisation requirements
separately in cosmologies beyond \LCDM{} to ensure unbiased dark-energy constraints.
\end{abstract}

\begin{keywords}
    {Galaxy clustering, weak gravitational lensing, cosmological parameters, dark energy}
\end{keywords}

\maketitle

\section{Introduction}\label{sec:intro}

The concordance cosmological model, \LCDM{}, describes a spatially flat Universe whose present-day
energy budget is dominated by cold dark matter and a cosmological constant, and it reproduces a
remarkable range of independent observations, from the cosmic microwave background
\citep{refId0} to the clustering of galaxies on large scales \citep[e.g.,][]{DES_Y3}, and the
distance--redshift relation of type Ia supernovae \citep[e.g.,][]{Jones_2022}. Its success notwithstanding,
the physical nature of the component driving the observed late-time acceleration, referred to as dark energy, remains unknown.
Within general relativity, a cosmological constant requires an equation-of-state (EoS) parameter
$w \equiv p/(\rho c^2) = -1$ at all times, so any robust detection of $w \neq -1$ would signal a
departure from \LCDM{} \citep{frieman2008dark}. A dynamical dark-energy component is commonly
parameterised through the Chevallier--Polarski--Linder form \citep{CHEVALLIER_2001, Linder},
\begin{equation}\label{eq:cpl}
    w(z) = w_0 + w_a \frac{z}{1+z},
\end{equation}
where $w_0$ is the present-day value of the EoS parameter and $w_a$ encodes its evolution; we refer
to this extension as \wowa{}. Recent baryon acoustic oscillation measurements from DESI, combined
with CMB data, exhibit a 3.1$\sigma$ preference for $w_0 > -1$ and $w_a < 0$
\citep{DESI}, a result that is sensitive to priors and to parameter degeneracies, and that
therefore places a high premium on the control of systematics in the independent late-time probes
that will test it.

The most constraining of those late-time probes is the combination of cosmic shear, photometric
galaxy clustering, and galaxy--galaxy lensing, commonly referred to as a \txtp{} analysis \citep[e.g.,][]{van_Uitert, DES_Y3}. The
three two-point functions carry complementary information: galaxy clustering traces the
inhomogeneous matter distribution modulated by galaxy bias, cosmic shear is sensitive to the
line-of-sight-integrated gravitational potential, and galaxy--galaxy lensing bridges the two by
correlating galaxy positions with the surrounding matter field. Their combination breaks
degeneracies, such as the degeneracy between $\Omega_{\rm m}$ and $\sigma_8$, and enables the self-calibration of several systematic effects
\citep{tutusaus2020euclid}. Stage-III surveys such as the Dark Energy Survey \citep[DES,][]{Abbott_2018, Abbott_2022} and
the Kilo Degree Survey \citep[KiDS,][]{Asgari-kids1000, kids_legacy} have demonstrated the power of this combination, reaching
percent-level precision on the growth of structure. The Stage-IV surveys now underway, including
\textit{Euclid} \citep{laureijs2011euclid, EuclidSkyOverview}, the Vera C.\ Rubin Observatory's Legacy
Survey of Space and Time \citep[LSST,][]{LSST_rubin}, and the Nancy Grace Roman Space Telescope's High-Latitude Wide-Area Survey 
\citep[HLWAS,][]{akeson2019widefieldinfraredsurvey}, will reduce the statistical uncertainties by a further
large factor.

To map the three-dimensional distribution of galaxies, \textit{Euclid} will estimate the redshift of each galaxy from its broadband photometric measurements, yielding a photometric redshift. Although less precise than spectroscopic redshifts, photometric redshifts can be obtained for much larger galaxy samples, making them essential for large photometric surveys. As photometric redshifts are estimated from broadband fluxes, rather than from spectral
features, the radial information in a photometric survey is degraded, and the analysis proceeds
tomographically. Galaxies are assigned to redshift bins, six bins for \textit{Euclid} Data Release 1 (DR1), and the observables are the angular power
spectra of, and between, those bins \citep{Hu_1999}. Every projection kernel entering the theory
prediction is a function of the redshift distribution of the corresponding bin, $n_i(z)$, so the
accuracy of the inferred cosmology is limited by the accuracy to which $n_i(z)$ is known. For
\textit{Euclid}, the source and lens distributions will be estimated primarily by mapping galaxies
into a multi-dimensional colour space with a self-organising map anchored to a large external
spectroscopic sample \citep{Masters_2015, EuclidSkyOverview}. These distributions are then subsequently calibrated using complex image simulations designed to reproduce the full complexity of the \textit{Euclid} scientific analysis pipeline. Nonetheless, the estimation and calibration process is imperfect, and thus residual uncertainties remain. These uncertainties must be
marginalised over during inference.

In practice this marginalisation is almost always performed with a low-dimensional parametric
model. The most widespread choice allows the mean of each tomographic bin to shift,
$n^i(z) \rightarrow n^i(z - \Delta z^i)$, introducing one nuisance parameter per bin
\citep{hamana2020cosmological, secco2022dark, PhysRevD.108.123519}. This is well motivated for
cosmic shear, whose signal is dominated by the mean redshift of the sources \citep{Ma_2006}, but
the photometric clustering signal responds to the full shape of the distribution \citep{cawthon},
so a shift-only treatment is not obviously sufficient for a \txtp{} analysis. Adding a per-bin
width rescaling, the so-called shifts-and-widths (S\&W) model, captures the next moment, yet forecasts for
Stage-IV precision indicate that even this is too rigid
\citep{ruizzapatero2026propagatingdatadrivengalaxyredshift}. This has motivated a range of more
flexible approaches: Gaussian mixture models \citep{nzmodel}, probability integral transforms
\citep{Myles}, explicit outlier parameterisations
\citep{zhang2025forecastingimpactsourcegalaxy}, the low-dimensional mapping of a discrete ensemble
of plausible realisations used in DES~Y3 \citep{cordero_DES}, the mode-projection scheme adopted
for the DES~Y6 \txtp{} analysis \citep{bernstein2026dimensionalreductionsampledpriors,
yin2026darkenergysurveyyear, giannini2025darkenergysurveyyear}, and the principal-component
marginalisation proposed for LSST \citep{ruizzapatero2026propagatingdatadrivengalaxyredshift}.

A feature shared by essentially all of these methods is that they have been developed and validated
in a \LCDM{} cosmology, and then applied unchanged in analyses whose stated goal is to constrain
dark energy. Whether that transfer is safe is not obvious: $w_0$ and $w_a$ modify both the
expansion history and the growth of structure, and can therefore compensate, or be compensated
by, changes in the projection kernels induced by a deformation of \nz{}. Establishing whether, and
how strongly, the choice of \nz{} uncertainty model imprints itself on the inferred dark-energy
parameters requires an \nz{} model flexible enough to explore deformations well beyond a shift and
a stretch, and cheap enough to be evaluated inside the likelihood.

In this work we present such a model. We reconstruct the redshift distribution of each tomographic
bin on the fly with a Gaussian Process \citep[GP,][]{rasmussen_GP}, conditioned on a small
number of training nodes whose positions and amplitudes are sampled jointly with the cosmological
and nuisance parameters. GPs define a prior over functions rather
than over the coefficients of a fixed functional form, they encode smoothness through an
interpretable correlation length, and their predictive mean is available in closed form, so the
reconstruction costs little relative to the Boltzmann and power-spectrum calls that dominate the
likelihood evaluation. We enforce positivity by construction, by performing the regression on
$\sqrt{n(z)}$, and we work in a logarithmic redshift coordinate so that the effective resolution of
the reconstruction follows the narrowing of the low-redshift bins.

Our framework is built around three pillars: (i) accurate modelling of the tomographic angular
power spectra of the photometric probes ($C_\ell^{\rm GG}$, $C_\ell^{\rm GL}$, $C_\ell^{\rm LL}$),
(ii) non-parametric, data-driven reconstruction of \nz{} via GPs with physically motivated
constraints, and (iii) Bayesian inference that propagates all relevant systematic uncertainties
through the public \texttt{cloe-org} ecosystem developed for \textit{Euclid} and Stage-IV cosmology. We apply our method to synthetic
\textit{Euclid}-DR1-like \txtp{} data vectors and compare a large suite of runs spanning two
cosmological models (\LCDM{} and \wowa{}), two probe combinations (weak lensing only and the full
\txtp{}), and a range of \nz{} uncertainty models (S\&W and GP reconstructions with
three to five training nodes) with two different sets of prior distributions imposed on GP hyperparameters (empirically informed and ensemble based). 

The paper is organised as follows. Section~\ref{sec:GP} details the GP formulation for
reconstructing \nz{}, including the coordinate transformation, the positivity and normalisation
constraints, the role of the number of training nodes and of the correlation length, and the
hyperparameter priors. Section~\ref{sec:methodology} describes the theoretical modelling of the
tomographic observables, the construction of the synthetic data vectors and covariance, and the
Bayesian inference set-up, together with the metrics we use to compare runs.
Section \ref{sec:results} and \ref{subsec:flows} present our results for the empirical and ensemble-based priors, respectively, focusing on the bias and constraining power of the
different \nz{} uncertainty models and on how both change when the dark-energy sector is opened.
Finally, Section~\ref{sec:conclusions} summarises our findings.

\section{Modelling galaxy redshift distributions using Gaussian processes}\label{sec:GP}

In this section we describe how the redshift distributions are reconstructed inside the
likelihood. We first summarise the GP formalism (Section~\ref{subsec:gp_formalism}), then specify
the reconstruction scheme and the transformations that guarantee physically admissible
distributions (Section~\ref{subsec:nz}), discuss the two hyperparameters that control the
flexibility of the model (i.e., the number of training nodes and the correlation length), and finally
define the prior distributions adopted for them (Section~\ref{subsec:gp_priors}).

Throughout, the uncertainty in \nz{} is parameterised through the training nodes on which the GP is
conditioned. Their positions and amplitudes, together with a correlation length per sample, are the
quantities sampled during inference, and the reconstruction they generate is what enters the
projection kernels. The uncertainty region in \nz{} is therefore not specified in advance but is
whatever the prior on those hyperparameters induces in function space. This flexibility is both the strength
of the approach and, as Section~\ref{sec:results} shows, the place where it currently needs the
most care. The reconstruction is publicly implemented in the Python package
\texttt{binz}\footnote{\url{https://github.com/jipdebuck/binz}}.

\subsection{Gaussian-process formalism}\label{subsec:gp_formalism}

As Stage-IV surveys seek to model increasingly flexible uncertainties in $n(z)$, GPs provide a natural choice because of their flexible, non-parametric nature. A GP defines a probability distribution over functions such that the function values at any finite set of input locations follow a joint Gaussian distribution, fully specified by its mean vector and covariance matrix. The input
locations at which the process is conditioned are the training nodes $\{\mathbf{x},\mathbf{y}\}$.
 Following common practice
in astronomical applications we set the mean function to zero \citep{GPastro}, so that the
structure of the process is entirely encoded by the kernel. To build the covariance matrix, we adopt a radial basis function
(RBF, or squared-exponential) kernel multiplied by a constant amplitude,
\begin{equation}\label{eq:rbf}
    k(x_i, x_j, l) = \sigma_{\rm amp}^2
    \exp\!\left[-\frac{d(x_i,x_j)^2}{2 l^2}\right],
\end{equation}
where $d(x_i,x_j)$ is the Euclidean distance between the two locations and $l$ the correlation length,
which sets the separation over which function values remain correlated. The RBF kernel is a natural
choice for this problem, as it captures the expected smooth variations of \nz{} with redshift.

Given the training set, the predictive mean of the process at a dense set of evaluation points
$\mathbf{x}^\ast$ is available in closed form,
\begin{equation}\label{eq:gp_pred}
    \hat{y}(\mathbf{x}^\ast) = k(\mathbf{x}^\ast, \mathbf{x})\, K^{-1}\, \mathbf{y},
\end{equation}
and the corresponding predictive variance quantifies the uncertainty induced by the GP prior
between nodes. Equation~\eqref{eq:gp_pred} is the only operation performed inside the likelihood,
which makes the reconstruction inexpensive. $K$ is of the size of the number of training nodes per
bin; that is, three, four or five.

\subsection{Reconstruction of the redshift distribution}\label{subsec:nz}

We model the redshift distribution in each of the $N_{\rm bin}$ tomographic bins as an independent
GP in redshift space, with a zero mean function, an RBF kernel, and a single correlation length
shared across bins. Sharing $l$ across bins keeps the number of
nuisance parameters, and hence the cost of the nested-sampling run, as low as possible while
retaining the flexibility that motivates the method in the first place. We use a separate
correlation length for the source and the lens samples, $l_{\rm she}$ and $l_{\rm pos}$.

\subsubsection{Transformed redshift coordinate.}
Photometric redshift uncertainties grow with redshift, so the low-redshift bins are narrower and
show more rapid variation than the high-redshift ones. To distribute the effective resolution of
the reconstruction accordingly, and to improve numerical stability at high redshift, we perform the
regression in the transformed coordinate
\begin{equation}\label{eq:logz}
    x = \frac{\log(1+z)}{\log(1+z_{\rm max})},
\end{equation}
which maps $[0, z_{\rm max}]$ onto $[0,1]$ and yields a finer resolution at low redshift. We adopt
$z_{\rm min} = 0$ and $z_{\rm max} = 3$ throughout.

\subsubsection{Positivity and normalisation.}
A normalised redshift distribution is a probability density and must therefore be strictly
positive, a constraint a GP does not respect by default. We enforce it by performing the regression on
\begin{equation}\label{eq:sqrt}
    y(z) = \sqrt{n(z)},
\end{equation}
so that the reconstruction $n(z) = y^2(z)$ is positive by construction. The resulting distribution
is renormalised to unit integral over $[z_{\rm min}, z_{\rm max}]$ before it is passed to the
projection kernels. 

\paragraph{Boundary and interior training nodes.}
In each bin, the first two training nodes are fixed at the edges of the redshift range,
\begin{align}
    x_{\rm min} &= \frac{\log(1+z_{\rm min})}{\log(1+z_{\rm max})} = 0, \quad y_{\rm min} = 0, \\
    x_{\rm max} &= \frac{\log(1+z_{\rm max})}{\log(1+z_{\rm max})} = 1, \quad y_{\rm max} = 0,
\end{align}
which forces $n_i(z)$ to vanish at both ends and removes physically implausible reconstructions.
The remaining, interior nodes are free: their positions $x_{i,j}$ and amplitudes $y_{i,j}$ are
sampled during inference. 

\subsubsection{Number of training nodes}\label{subsubsec:nodes}

The number of interior nodes per bin controls how much freedom the GP has, and sets the
dimensionality of the \nz{} sector of the parameter space. With a single node, the process has one
constraint per bin and essentially no information about the shape of the distribution, so the
reconstruction is dominated by the shape implied by the kernel, that is a Gaussian for the RBF choice. The single-node reconstruction is therefore a
decent approximation in the central Gaussian-like bins, but it fails in the first and the last bin, where the
true distributions are markedly skewed. Sampling the position of a single node therefore captures
the uncertainty in the mean redshift, much as a shift parameter does, but not higher-order
deformations of the tails or the widths.

\begin{figure}[h!]
    \centering
    \includegraphics[width=\linewidth]{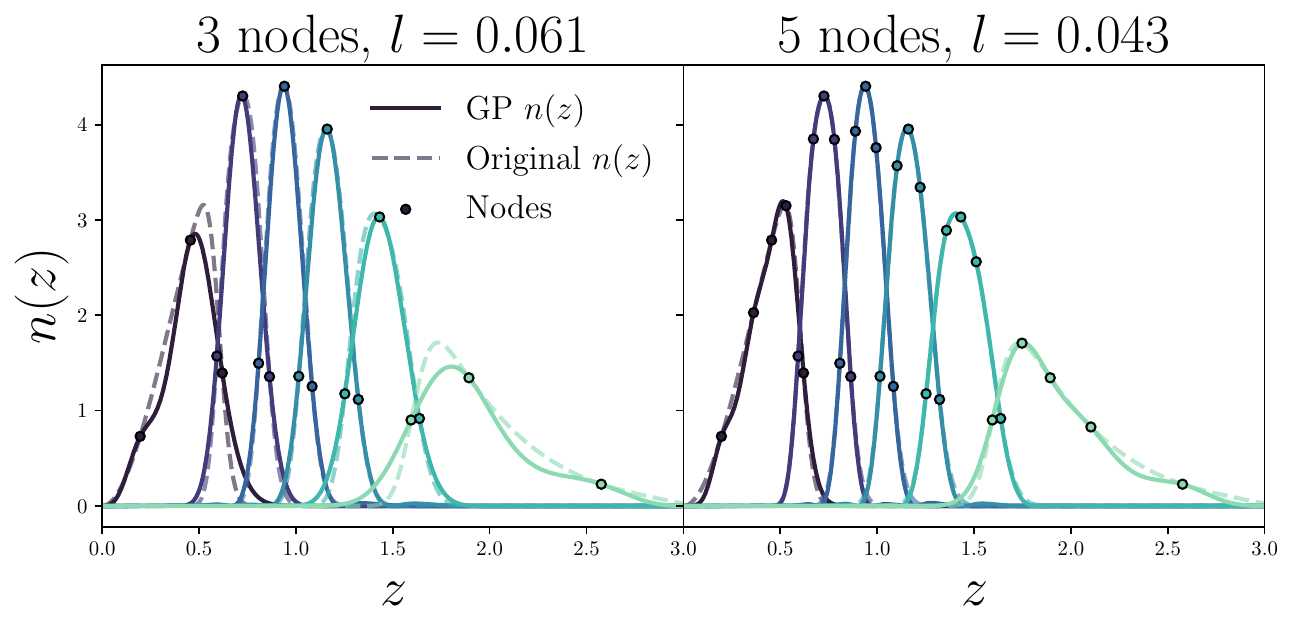}
    \caption{Unnormalised GP reconstructions with three (left) and five (right) interior training
    nodes per bin, with correlation lengths $l = 0.061$ and $l = 0.043$ respectively. More nodes
    give the GP the freedom to depart from the Gaussian shape implied by the RBF kernel, at the
    cost of a larger nuisance-parameter space.}
    \label{fig:nodes_comparison}
\end{figure}

Adding nodes supplies additional constraints and reduces the influence of the kernel on the overall
shape, allowing asymmetric, skewed and non-Gaussian distributions to be reproduced.
Figure~\ref{fig:nodes_comparison} compares reconstructions with three and five interior nodes. The
three-node configuration still struggles in the outermost bins, whereas the five-node configuration
recovers their shape more accurately. The price is a larger parameter space, a higher computational
cost, and an increased risk of the model accommodating features that the data do not require.

\subsubsection{Correlation length}\label{subsubsec:corrlen}

The correlation length sets the smoothing scale of the reconstruction. A value that is too small permits
unconstrained oscillations between adjacent nodes, whereas a value that is too large produces oversmoothed and
excessively broad distributions. Crucially, its interpretation depends on the node density.
With a single node, $l$ mainly controls the width of
the reconstructed distribution, whereas with three or five nodes it controls the distance over
which neighbouring nodes still influence one another. The node spacing decreases as the number of nodes increases, resulting in a smaller correlation length for a larger number of nodes (see Fig. \ref{fig:nodes_comparison}). 
The priors on the correlation length are defined accordingly in Section~\ref{subsec:gp_priors}.

The parameters varied during cosmological inference are the per-bin $x$- and
$y$-positions of the interior training nodes, plus a single global correlation length per sample. All other GP inputs are kept fixed: the kernel is RBF with a constant amplitude
$\sigma_{\rm amp}^2 = 0.04$, no observational noise is added to the training nodes, and the mean
function is zero. We consider cases with three, four, and five interior training nodes, since configurations with one or two training nodes are not sufficient to reconstruct the full shape of the redshift distribution. 

\subsection{Priors on the GP hyperparameters}\label{subsec:gp_priors}

A Bayesian analysis requires priors for every sampled parameter, and for a non-parametric model the
prior is where the calibration enters. We consider empirical and ensemble-based prescriptions, applied to
different GP configurations and summarised in Table~\ref{tab:gp_priors}.  

\subsubsection{Empirically informed priors} \label{emp_priors}
In the empirical approach, the fiducial node positions are defined by quantiles of the calibrated distribution together with its peak: at
the $0.05$ and $0.95$ quantiles and the peak for three nodes; at the $0.05$, $0.35$, $0.65$ and
$0.95$ quantiles for four nodes; and at the $0.05$, $0.275$, $0.5$, $0.725$ and $0.95$
quantiles for five nodes. The prior widths are then set empirically, by drawing samples from
    the prior before inference and checking that the resulting spread of reconstructions is
    consistent with a plausible ensemble of redshift distributions
    (Fig.~\ref{fig:samples_GP}). This procedure yields a uniform spread of $\pm 0.05$ around each
    fiducial node position for all multi-node cases. The correlation-length priors shift to
    progressively smaller values as the number of nodes increases, following
    Section~\ref{subsubsec:corrlen}. The empirical priors adopted for the different node configurations are given in Table \ref{tab:gp_priors}
    
The middle panel of Fig. \ref{fig:samples_GP} shows 20 samples drawn from the empirical prior in the three-node configuration. We stress that the empirical prior is deliberately conservative. As evident from Fig. \ref{fig:samples_GP}, it reflects a large uncertainty in
the calibrated \nz{} in order to give the GP sufficient freedom. As we show in
Section~\ref{sec:results}, this conservatism is also the principal limitation of the current
implementation, and motivates the ensemble-calibrated priors discussed below.

\subsubsection{Ensemble-based priors}\label{ensemble_priors}
The \textit{Euclid} source and lens photometric redshift distributions are estimated using a large external sample of spectroscopic redshifts \citep{EuclidSkyOverview}. The
subsequent $n(z)$ calibration procedure results in an ensemble of plausible redshift distributions that capture the statistical and systematic uncertainties in the photometric redshift estimates. By constructing a prior based on the ensemble, we can propagate $n(z)$ uncertainties in a data-driven manner.

We construct this data-driven prior using a normalising flow \citep{norm_flow}: a generative model that represents
a complex probability distribution as an invertible, differentiable transformation of a simple base
distribution, and that provides an explicit, tractable density. Given an ensemble of plausible
calibrated \nz{} realisations, the flow can be trained to
learn the joint distribution of the GP hyperparameters, that is of the node positions, node
amplitudes and correlation lengths that
reproduce that ensemble. Since we train a separate flow per bin, we adopt a bin-specific correlation length rather than a single global correlation length. A trained flow then supplies the prior probability of
the currently proposed set of GP hyperparameters during inference, so that the reconstruction is bounded by what the
photometric calibration actually supports rather than by a uniform box chosen by hand. Because
flows are differentiable by construction, this also keeps the door open to gradient-based samplers
for the full joint posterior.

\begin{figure*}[t]
    \centering
    \includegraphics[width=\textwidth]{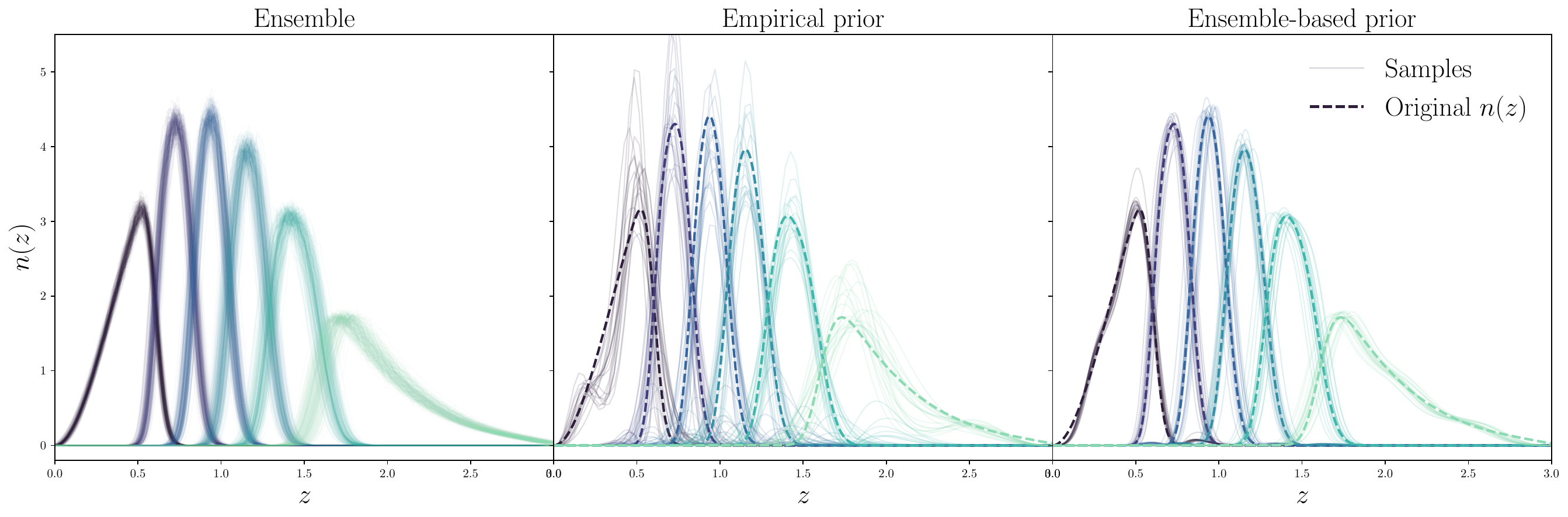}
    \caption{The left panel shows the ensemble of redshift distributions used to derive the ensemble-based prior. The middle and right panels show twenty samples drawn from the empirical and ensemble-based prior distributions in the three-node GP configuration, respectively. Following the empirical prior, the node positions are drawn from a uniform distribution with a spread of $\pm 0.05$ around the fiducial node positions, and the correlation length is sampled from a uniform distribution between 0.05 and 0.08. The ensemble-based prior is constructed by training a normalising flow on the GP hyperparameter distributions derived from the ensemble of plausible redshift distributions shown in the left panel.}
    \label{fig:samples_GP}
\end{figure*}

The left panel of Fig. \ref{fig:samples_GP} shows the ensemble of redshift distributions from which we derive the ensemble-based prior by training a normalising flow per bin. The resulting prior is shown in the right panel of Fig. \ref{fig:samples_GP}, where we draw 20 samples from the ensemble-based prior distribution in the three-node configuration. The ensemble-based prior is more tightly constrained than the empirical prior and better reflects the uncertainties represented by the ensemble, which was the primary objective of this approach.

\begin{table}[h!]
\centering
\renewcommand{\arraystretch}{1.8}
\setlength{\tabcolsep}{19pt}
\begin{tabular}{lc}
\hline\hline
\textbf{Parameter}  & \textbf{Empirical prior} \\
\hline
$x^{i \in [1,N_{\rm bin}]}_{j \in [1,5]}$ &
$\mathcal{U}(x^i_{j,\rm fid} \pm 0.05)$ \\
$y^{i \in [1,N_{\rm bin}]}_{j \in [1,5]}$ & $\mathcal{U}(y^i_{j,\rm fid} \pm 0.05)$ \\
$\ln l_{\rm 3\,nodes}$ &  $\mathcal{U}(\ln 0.05, \ln 0.08)$ \\
$\ln l_{\rm 4\,nodes}$ &  $\mathcal{U}(\ln 0.04, \ln 0.08)$ \\
$\ln l_{\rm 5\,nodes}$ &  $\mathcal{U}(\ln 0.02, \ln 0.06)$ \\
\hline
\end{tabular}
\caption{Empirical prior distributions adopted for the GP hyperparameters. $\mathcal{U}$ and $\mathcal{N}$
denote uniform and Gaussian distributions. For the node coordinates, $i$ indexes the tomographic
bin and $j$ the interior node within that bin.}
\label{tab:gp_priors}
\end{table}

\section{Methodology}\label{sec:methodology}

\subsection{Modelling of the photometric probes}\label{subsec:probes}

We model the tomographic angular power
spectrum between probes $A$ and $B$ measured in bins $i$ and $j$, $C^{AB}_{ij}(\ell)$, following the theoretical prescription in \cite{theoreticalrecipe}. Importantly, the radial weight functions necessary to compute $C^{AB}_{ij}(\ell)$ depend differently on the photometric redshift distribution across probes: the cosmic shear kernel depends primarily on the mean of the $n(z)$, whereas the galaxy clustering kernel is directly proportional to $n(z)$. The two probes therefore respond differently to changes in, for example, the $n(z)$ width \citep{Ma_2006, cawthon}.

The observed spectra include relevant astrophysical and observational systematics. For cosmic shear we model intrinsic alignments (IA) with the extended non-linear alignment prescription,
adding the shear--IA and IA--IA terms,
$C^{\rm LL}_{ij} = C^{\gamma\gamma}_{ij} + C^{\gamma \rm I}_{ij} + C^{\rm II}_{ij}$, parameterised
by an amplitude $\mathcal{A}_{\rm IA}$ and a redshift-evolution slope $\eta_{\rm IA}$
\citep{TROXEL20151}. For galaxy clustering we include magnification bias,
$C^{\rm GG}_{ij} = C^{\rm gg}_{ij} + C^{{\rm g}\mu}_{ij} + C^{\mu\mu}_{ij}$
\citep{magnification}, with one magnification-bias parameter per lens bin, and we parameterise the
linear galaxy bias as a cubic polynomial,
$b_{\rm G}(z) = b_{{\rm G},0} + b_{{\rm G},1} z + b_{{\rm G},2} z^2 + b_{{\rm G},3} z^3$.
Galaxy--galaxy lensing, $C^{\rm GL}_{ij}$, is computed analogously from the cross-correlation of
the clustering and lensing kernels. Shape-measurement uncertainties are captured by a multiplicative shear bias $m^i_{\rm L}$ nuisance parameter per source bin. 

Non-linear corrections to the matter power spectrum, including baryonic feedback, are computed
with \texttt{HMCode2020} \citep{mead2021hmcode}, which introduces the AGN feedback temperature,
$\log_{10}(T_{\rm AGN}/{\rm K})$, as an additional free parameter. To speed up the calculations, we use the \texttt{HMCode2020Emu} emulator, trained on
\texttt{CAMB} \citep{Lewis:1999bs}, while the background quantities are obtained from \texttt{CAMB} directly. To produce consistent theoretical predictions, we use \texttt{cloelib}\footnote{\label{footnote_cloe}\url{https://github.com/cloe-org/cloelib}}
\citep[version 0.11.dev28,][]{bonici2026cloelibflexiblepythonlibrary}, which also implements the
theoretical model for the photometric probes. This set-up follows that of \citet{Overview} and
\citet{canas2025euclid}. The python package \texttt{cloelib} is part of the \texttt{cloe-org}\footnote{\href{https://github.com/cloe-org/cloelib}{\url{https://github.com/cloe-org}}} broader ecosystem.

\subsection{Synthetic data and covariance}\label{subsec:data}
The noiseless synthetic data vectors representative of a Stage IV first cosmological data release, such as \textit{Euclid} photometric survey, are also generated using \texttt{cloelib}. The galaxy
sample is split into six tomographic bins, matching the binning foreseen for \textit{Euclid} DR1 (dashed line in Fig.~\ref{fig:nodes_comparison}), as shown in \citet{bertmann_nz_lens}. Following a simplified approach as in \citet{Overview}, the same distribution is used for the source and the lens samples, but the two are treated as independent in the analysis. The data vector consists of the auto- and cross-spectra of all bins for the three
probe combinations, binned into $32$ logarithmically spaced multipole bins over
$\ell \in [10, 1500]$. The fiducial cosmology and nuisance parameters are listed in
Table~\ref{tab:priors}.

The covariance matrix $\mathbf{C}$ is analytical and includes Gaussian and super-sample contributions, and is generated with \texttt{Spaceborne}\footnote{\href{https://github.com/davidesciotti/Spaceborne}{\url{https://github.com/davidesciotti/Spaceborne}}}. We held it fixed throughout the analysis.

\subsection{Inference set-up}\label{subsec:inference}

We perform Bayesian inference on the cosmological parameters
\begin{equation}
    \theta = \{\Omega_{\rm cdm},\, \Omega_{\rm b},\, \ln(10^{10}A_{\rm s}),\, n_{\rm s},\, H_0,\,
    w_0,\, w_a\},
\end{equation}
together with the nuisance parameters
\begin{equation}
    \phi = \big\{\mathcal{A}_{\rm IA},\, \eta_{\rm IA},\, \log_{10} T_{\rm AGN},\,
    b_{{\rm G},0\ldots3},\, b^i_{\rm mag},\, m^i_{\rm L}\big\},
\end{equation}
and the \nz{} sector, which is either the standard shift and width parameters,
$\{\Delta z^i_{\rm she},\, \Delta z^i_{\rm pos},\, \sigma^i_{\rm she},\, \sigma^i_{\rm pos}\}$,
or, for the GP runs with empirical priors, the hyperparameters
\begin{equation}
    \psi = \big\{(x_{i,j},\, y_{i,j}),\, l_{\rm she},\, l_{\rm pos}\big\},
\end{equation}
and with ensemble-based priors
\begin{equation}
    \psi = \big\{(x_{i,j},\, y_{i,j}),\, l^i_{\rm she},\, l^i_{\rm pos}\big\},
\end{equation}
where the correlation lengths are bin specific. In the baseline models the shift and width transformations are
\begin{align}
    n^i_{\rm shifted}(z) &= n^i(z - \Delta z^i), \\
    n^i_{\rm stretched}(z) &= n^i\!\left[\sigma^i_z (z - z^i_{\rm mean}) + z^i_{\rm mean}\right],
\end{align}
followed by renormalisation. We run both a shifts-only variant and S\&W variant.

The likelihood is Gaussian in the data vector,
\begin{equation}\label{eq:like}
    \mathcal{L}(\mathbf{d}|\theta,\phi,\psi) \propto
    \exp\!\Big[-\tfrac{1}{2}\,\big(\mathbf{d}-\mathbf{t}(\theta,\phi,\psi)\big)^{\!\top}
    \mathbf{C}^{-1} \big(\mathbf{d}-\mathbf{t}(\theta,\phi,\psi)\big)\Big],
\end{equation}
with $\mathbf{d}$ the synthetic data vector and $\mathbf{t}$ the theory prediction. It is evaluated
with likelihood package of the \texttt{cloe-org} ecosystem, \texttt{cloelike}\footnote{\url{https://github.com/cloe-org/cloelike}} \citep[version 0.2.2.dev22,][]{bonici2026cloelikepythonlibrarycosmological}. The GP reconstruction is provided by
\texttt{binz}.

\begin{table*}[t]
\centering
\renewcommand{\arraystretch}{1.25}
\setlength{\tabcolsep}{18.9pt}
\begin{tabular}{lcc|lcc}
\hline\hline
\textbf{Parameter} & \textbf{Fiducial} & \textbf{Prior} &
\textbf{Parameter} & \textbf{Fiducial} & \textbf{Prior} \\
\hline
\multicolumn{3}{c|}{Cosmology} & \multicolumn{3}{c}{Nuisance parameters} \\
\hline
$\Omega_{\rm b}$ & $0.049$ & $\mathcal{U}(0.01, 0.1)$ &
$\mathcal{A}_{\rm IA}$ & $0.16$ & $\mathcal{U}(-1, 1)$ \\
$\Omega_{\rm cdm}$ & $0.27$ & $\mathcal{U}(0.1, 0.8)$ &
$\eta_{\rm IA}$ & $1.66$ & $\mathcal{U}(-5, 5)$ \\
$\ln(10^{10} A_{\rm s})$ & $3.04$ & $\mathcal{U}(1.79, 3.91)$ &
$b_{{\rm G},0}$ & $1.33291$ & $\mathcal{U}(-3, 3)$ \\
$n_{\rm s}$ & $0.96$ & $\mathcal{U}(0.6, 1.2)$ &
$b_{{\rm G},1}$ & $-0.72414$ & $\mathcal{U}(-3, 3)$ \\
$H_0$ & $67$ & $\mathcal{U}(50, 90)$ &
$b_{{\rm G},2}$ & $1.01830$ & $\mathcal{U}(-3, 3)$ \\
$w_0$ & $-1$ & $\mathcal{U}(-3, -0.5)$ &
$b_{{\rm G},3}$ & $-0.14913$ & $\mathcal{U}(-3, 3)$ \\
$w_a$ & $0$ & $\mathcal{U}(-3, 3)$ &
$m^{i \in [1,6]}_{\rm L}$ & $0$ & $\mathcal{N}(0, 0.01)$ \\
 & & &
$b^{i \in [1,6]}_{\rm mag}$ & $0$ & $\mathcal{U}(-2, 2)$ \\
 & & &
$\Delta z^{i \in [1,6]}_{\rm she}$ & $0$ & $\mathcal{N}(0, 0.002(1+z^{\rm fid}_i))$ \\
 & & &
$\Delta z^{i \in [1,6]}_{\rm pos}$ & $0$ & $\mathcal{N}(0, 0.002(1+z^{\rm fid}_i))$ \\
 & & &
$\sigma^{i \in [1,6]}_{\rm she}$ & $1$ & $\mathcal{N}(1, 0.01)$ \\
 & & &
$\sigma^{i \in [1,6]}_{\rm pos}$ & $1$ & $\mathcal{N}(1, 0.01)$ \\
\hline
\end{tabular}
\caption{Fiducial values and prior distributions of the cosmological and nuisance parameters. The
GP hyperparameter priors are given separately in Table~\ref{tab:gp_priors}. The neutrino mass is
fixed to $m_\nu = 0.077$~eV and the curvature density to $\Omega_K = 0$. In \LCDM{} runs $w_0$ and
$w_a$ are fixed to their fiducial values. The priors imposed on the shift parameters are based on \textit{Euclid} Data Release 3 (DR3) requirements for the accuracy of the mean of the redshift distributions.}
\label{tab:priors}
\end{table*}

Sampling is carried out with the nested sampler \texttt{dynesty} \citep{dynesty}, using
multi-ellipsoid bounding and reflective slice sampling, with $n_{\rm live} = 50 \times N_{\rm dim}$
initial live points. A chain is considered converged once the remaining evidence satisfies
$\Delta \ln \mathcal{Z} < 0.01$ or an effective sample size of $10^4$ is reached. We compared
\texttt{dynesty} with \texttt{nautilus} \citep{Lange}, which trains neural networks to propose live
points in regions of high likelihood. \texttt{nautilus} converges faster in general, but
\texttt{dynesty} proved more reliable for the multi-modal posteriors encountered in the more
flexible GP configurations, and we therefore adopt it throughout. 
The inference proceeds as follows:
\begin{enumerate}
    \item \textbf{Initialisation.} Define the cosmological, nuisance and \nz{} parameter spaces and
    their priors; build the synthetic data vector and the covariance; initialise the sampler.
    \item \textbf{\nz{} reconstruction.} For each proposed point, reconstruct $n_i(z)$ in every
    tomographic bin from the sampled hyperparameters $\psi$ via Eq.~\eqref{eq:gp_pred}, apply the
    positivity map of Eq.~\eqref{eq:sqrt}, and renormalise.
    \item \textbf{Theory computation.} Evaluate the tomographic angular power spectra with
    \texttt{cloelib} using the reconstructed $n_i(z)$ in all projection kernels.
    \item \textbf{Likelihood evaluation.} Evaluate Eq.~\eqref{eq:like} with \texttt{cloelike}, and
    let the sampler replace the lowest-likelihood live points.
\end{enumerate}
Steps 2--4 are repeated until convergence. Runs were parallelised on the LEONARDO supercomputer
hosted by CINECA \citep{turisini2023leonardo} under the \textit{Euclid} EuroHPC Extreme Scale
allocation.

\subsection{Metrics}\label{subsec:metrics}

We compare runs using three quantities. The bias $b_a$ induced by a given model on parameter $a$
is the offset of the marginalised posterior mean from the input value in units of the posterior
width,
\begin{equation}\label{eq:bias}
    b_a = \frac{\mu_a - a_{\rm fid}}{\sigma_a},
\end{equation}
and the total bias accumulated by a chain is $B = \big(\frac{1}{N}\sum_i b_i^2\big)^{1/2}$, where $N$ is the number of parameters. The
constraining power is quantified by the figure of merit
\begin{equation}\label{eq:fom}
    \mathrm{FoM}(a,b) = \frac{1}{\sqrt{\det \mathrm{Cov}(a,b)}},
\end{equation}
evaluated for $(w_0, w_a)$ in \wowa{} runs (i.e. the quantity \textit{Euclid} is designed to
maximise) and for $(\Omega_{\rm m}, S_8)$ otherwise, with
$S_8 = \sigma_8 \sqrt{\Omega_{\rm m}/0.3}$. Finally, models are compared through the Bayes factor
$\ln B_{i,{\rm ref}} = \ln \mathcal{Z}_i - \ln \mathcal{Z}_{\rm ref}$, which penalises the enlarged
prior volume of the more flexible models automatically.

We note an important implication of this inference process: as our data vectors are noiseless and generated from the
fiducial \nz{}, Eq.~\eqref{eq:bias} measures the bias induced purely by the \nz{} model and the
prior volume it opens, not by a statistical fluctuation. A model with a larger $B$ is
therefore genuinely mis-propagating the redshift information, and a model with both a large FoM and
a large $B$ is concentrating the posterior on the wrong region of parameter space.

\section{Reconstruction with empirical priors}\label{sec:results}

We now present the comparison of \nz{} uncertainty models when adopting empirical priors for GP hyperparameters (see Section \ref{emp_priors}). Rather than reproducing the full suite of
runs, we focus here on the result that is most consequential for Stage-IV dark-energy science: the
dependence of the inferred parameters on the adopted \nz{} model is substantially stronger in
\wowa{} than in \LCDM{}, in a way that no \LCDM{}-only validation would reveal.

\subsection{Impact of \nz{} model across different cosmologies}\label{subsec:w0wa}

\begin{figure*}[t]
    \centering

    \begin{subfigure}{0.48\textwidth}
        \centering
        \includegraphics[width=\linewidth]{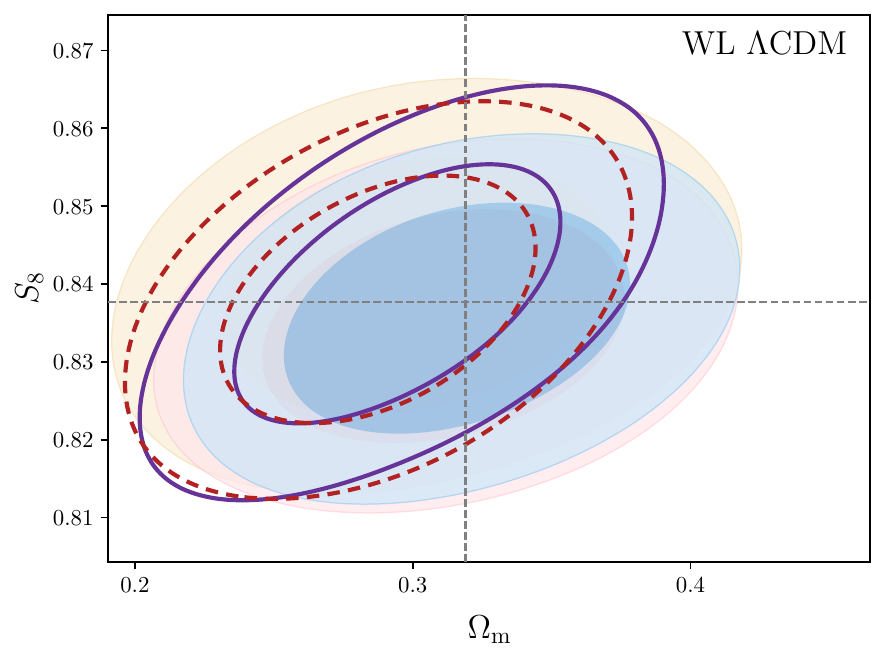}
    \end{subfigure}
    \hfill
    \begin{subfigure}{0.48\textwidth}
        \centering
        \includegraphics[width=\linewidth]{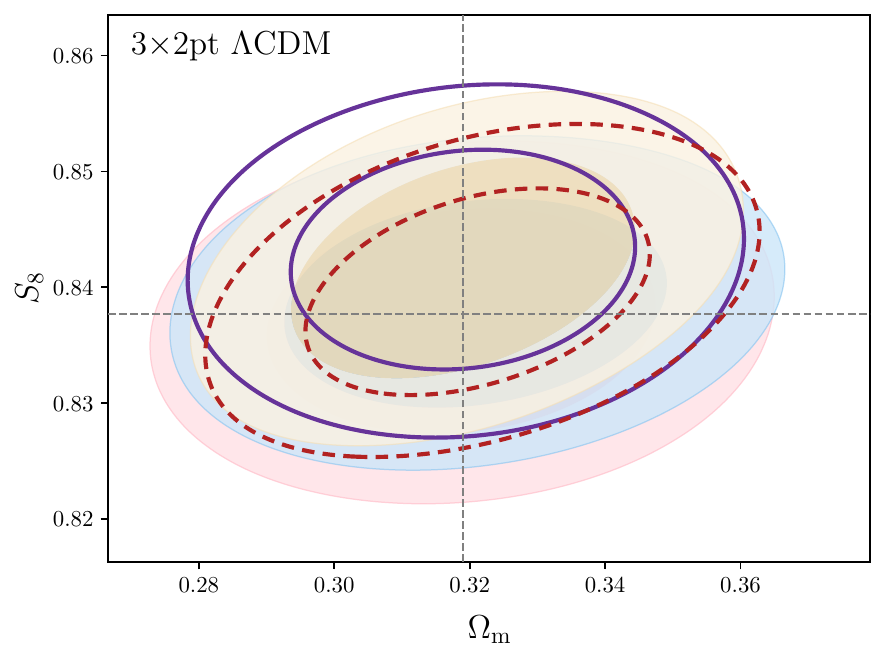}
    \end{subfigure}

    \vspace{0.5em}

    \begin{subfigure}{0.48\textwidth}
        \centering
        \includegraphics[width=\linewidth]{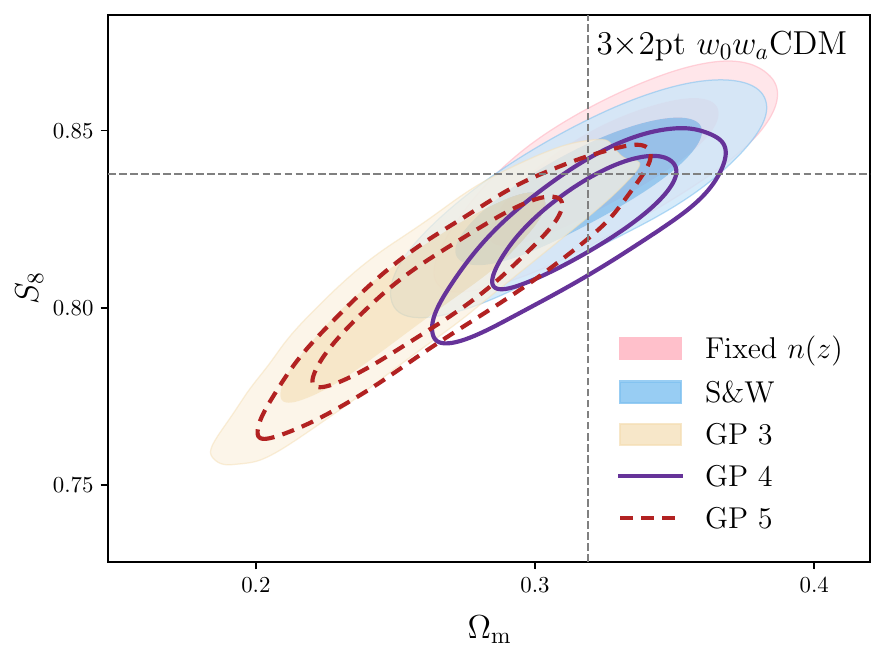}
    \end{subfigure}
    \hfill
    \begin{subfigure}{0.48\textwidth}
        \centering
        \includegraphics[width=\linewidth]{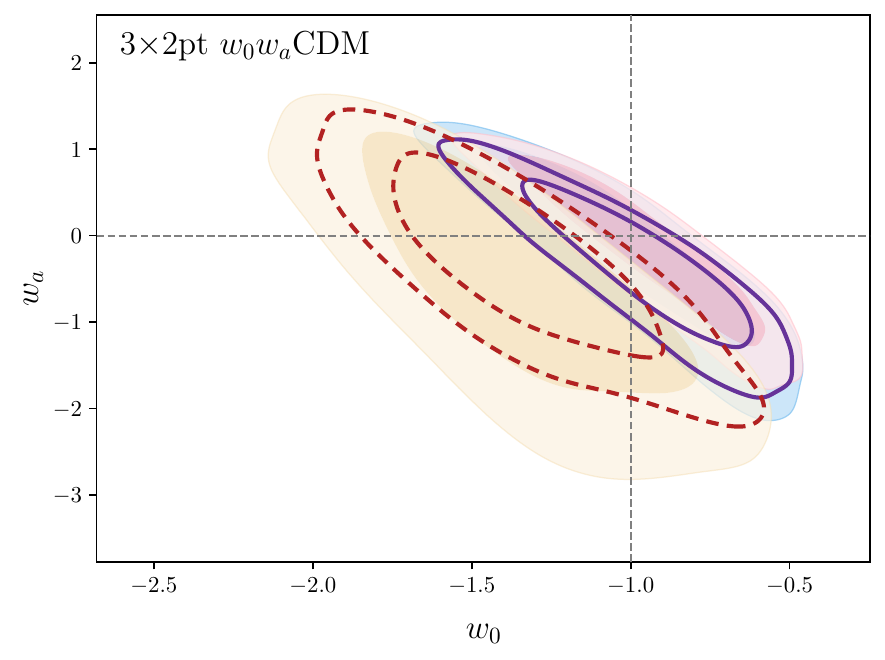}
    \end{subfigure}

    \caption{Marginalised two-dimensional ($\Omega_{\rm m}$-$S_8$) (top and bottom left) and ($w_0$-$w_a$) (bottom right) posteriors from WL \LCDM{} (top left), \txtp{} \LCDM{} (top right), and \txtp{} \wowa{} (bottom) chains with three, four and five
    interior training nodes. The S\&W and fixed-$n(z)$ runs are shown for reference and dashed grey lines mark the
    input values. Compared to \txtp{} \LCDM{}, the \txtp{} \wowa{} cases vary substantially across redshift uncertainty models, with the 4-node configuration tracking S\&W and the 3- and 5-node models closely following each other. The \txtp{} \wowa{} chains show a similar trend in the marginalised posterior of $w_0$ and $w_a$. We do not show WL \wowa{} runs, since cosmic shear leaves evolving dark energy largely unconstrained. Note that the axis limits differ between panels.  
    }
    \label{fig:omega_S8_nodes}
\end{figure*}

The top and bottom left panels of Fig.~\ref{fig:omega_S8_nodes} show the marginalised two-dimensional posteriors of
$\Omega_{\rm m}$ and $S_8$ from weak lensing (WL, i.e. cosmic shear only) \LCDM{}, \txtp{} \LCDM{} and \wowa{} chains, with three, four and five interior training
nodes, together with the S\&W and fixed-\nz{} reference. We refrain from showing WL \wowa{} runs, since the constraining power of cosmic shear in \wowa{} is limited. The WL \LCDM{} chains do constrain $\Omega_\text{m}$ and $S_8$ (top left panel), although the GP models introduce a slight bias in $\Omega_\text{m}$. 

The inclusion of galaxy clustering and galaxy-galaxy lensing increases the constraining power, allowing us to compare the results across cosmological models. The ($\Omega_{\rm m}$--$S_8$) posteriors in \wowa{} (bottom left panel) display a pattern where the four-node
contours remain relatively consistent with the S\&W and fixed-\nz{} cases and cluster around the input cosmology, while the
three- and five-node contours are noticeably displaced from it. The ($\Omega_{\rm m}$--$S_8$)
posteriors of the latter two shift along the degeneracy direction towards lower values of both
parameters. The corresponding
\LCDM{} runs (top right panel), in contrast, agree significantly better to within a small
fraction of the posterior width. Hence, the additional freedom introduced by $w_0$ and $w_a$ is what
amplifies the impact of the \nz{} reconstruction on the inferred cosmology, and the effect of \nz{} uncertainties on the posterior distributions of $\Omega_\text{m}$ and $S_8$ depend on the adopted cosmological model.

The bottom right panel of Fig. \ref{fig:omega_S8_nodes} shows the two-dimensional marginalised posteriors of $w_0$ and $w_a$ from \txtp{} chains. The \txtp{} ($w_0$-$w_a$) posteriors show a similar trend as ($\Omega_\text{m}$-$S_8$), where the 3-node model tracks the 5-node model, and the 4-node configuration follows the S\&W and fixed-\nz{} posteriors. The similarity in the posteriors of the two parameter combinations highlights that deviations in \nz{} affect not only the inferred dark-energy parameters, but also the interpretation of $\Omega_\text{m}$ and $S_8$ through their degeneracy with $w_0$ and $w_a$. 

The Bayesian evidences confirm that none of the GP configurations is preferred by the data in
\wowa{}: we obtain $\ln \mathcal{Z} \approx -78.6$ for the three-node model, $-137.0$ for the four-node model and $-78.5$ for the five-node model, against $\ln \mathcal{Z} \approx -59.0$ for S\&W, $-57.4$ for shifts-only and $-54.4$ for a fixed \nz{}. The
four-node model is disfavoured by a large margin, which is at first sight surprising given that its
cosmological constraints are the closest to the reference. To explore this, we investigate the \nz{} reconstructions preferred by the data in Appendix \ref{Appendix_C}. The preferred 4-node \nz{} reconstruction deviates significantly from the fiducial \nz{} (see Fig. \ref{fig:preferred_nz_w0wa}), with the resulting bias absorbed primarily by shifts in the nuisance rather than cosmological parameters, explaining the small evidence of the 4-node chain despite its unbiased cosmological constraints. The departure from the input \nz{} is primarily driven by the small preferred correlation length in the \wowa{} 4-node configuration (see Fig. \ref{fig:hyper}). The absence of a similar effect in the \LCDM{} counterpart suggests that the inclusion of the dark-energy parameters causes the shift in the preferred \nz{}. Based on the evidence, the data prefer a fixed-\nz{}. This is expected, as the noiseless data vectors are generated using the fiducial \nz{}, meaning that a model that introduces the fewest deviations from this fiducial distribution will naturally be preferred.

The overall impact of opening up the dark energy sector becomes clear when considering the biases across the parameter space (see Appendix \ref{appendix_A}), together with the constraining power of the different \nz{} models (see Appendix \ref{Appendix_B}). Taken together, these results show that biases are larger in a \wowa{} cosmology, the preferred \nz{} model depends on the adopted cosmological model, and tight constraints do not guarantee unbiased parameter estimates. Redshift distribution uncertainty methods should thus be evaluated separately when considering extensions of \LCDM{}, and constraining power estimates should be accompanied by a bias diagnostic.

\subsection{Implications for Stage-IV dark-energy analyses}\label{subsec:implications}

The results above have a direct consequence for how \nz{} uncertainty models should be validated.
The methods deployed by Stage-III surveys, such as the per-bin shifts in DES~Y1, shifts plus lens widths in
DES~Y3, the low-dimensional ensemble mapping of \citet{cordero_DES}, and the mode-projection scheme
of DES~Y6 \citep{bernstein2026dimensionalreductionsampledpriors}, were all calibrated in a \LCDM{}
cosmology and subsequently applied unchanged to constrain dark energy
\citep{descollaboration2026constraintsdynamicaldarkenergy}. Our findings indicate that this
transfer is not automatically robust. Not only does the sensitivity of the constraints to the \nz{}
treatment grow when $w_0$ and $w_a$ are freed, the \nz{} shape that the data prefer changes as
well. This matters most for the
methods that are designed to constrain \nz{} beyond its first two moments. Mode projection selects
the directions of \nz{} variation with the largest cosmological impact, and the PCA marginalisation
proposed for LSST \citep{ruizzapatero2026propagatingdatadrivengalaxyredshift} defines its basis from
the residuals of an ensemble. In both cases the basis is implicitly conditioned on the cosmological
model used to construct it. If the dominant directions of variation are cosmology dependent, a
basis derived in \LCDM{} may be an inefficient, or biased, description of the uncertainty in
\wowa{}. A comparison of \nz{} uncertainty models carried out directly in \wowa{} therefore seems
necessary to ensure robust dark-energy constraints.

\begin{figure*}[t]
    \centering

    \begin{subfigure}{0.48\textwidth}
        \centering
        \includegraphics[width=\linewidth]{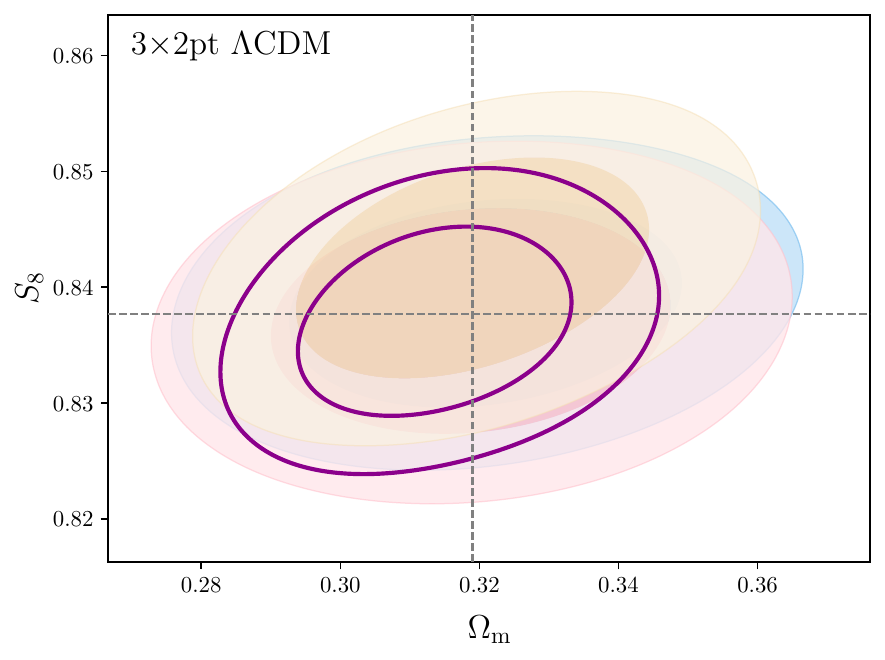}
    \end{subfigure}
    \begin{subfigure}{0.48\textwidth}
        \centering
        \includegraphics[width=\linewidth]{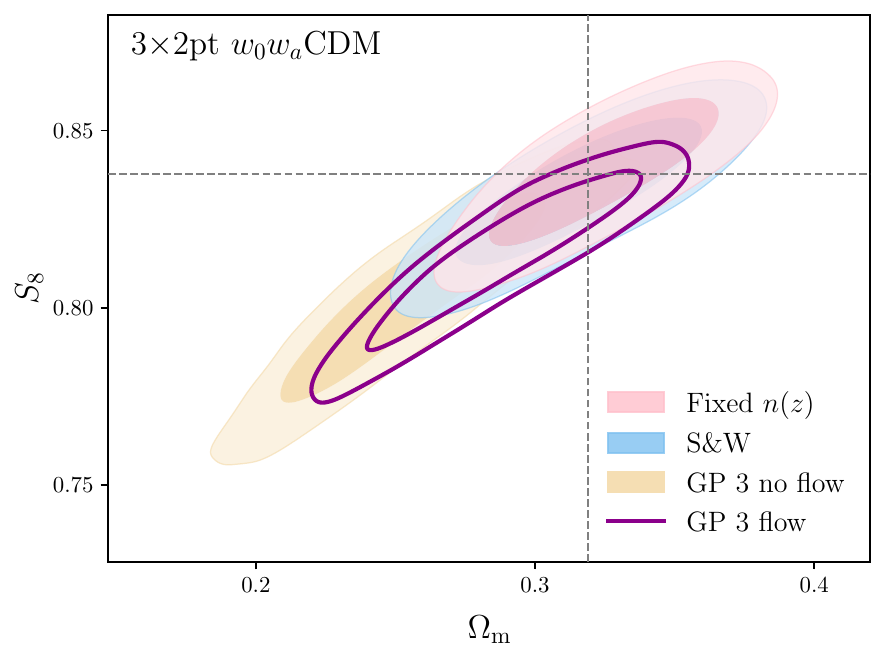}
    \end{subfigure}
    
    \caption{Marginalised two-dimensional ($\Omega_{\rm m}$-$S_8$) posteriors from \txtp{} chains in \LCDM{} (left) and \wowa{} (right) cosmologies. The flow (i.e., ensemble-based priors) and no-flow (i.e., empirical priors) posteriors are shown as the unfilled purple and filled yellow contours, respectively. The S\&W baseline case is shown for reference, and dashed grey lines represent the input values. In \LCDM{}, the ensemble-based prior reduces the posterior area with respect to the empirical and S\&W prior. In \wowa{}, although still present, the discrepancy between S\&W and GP cases decreases when adopting the ensemble-based prior. Note that the axis limits differ between panels.
    }
    \label{fig:flow_posteriors}
\end{figure*}

\cite{bertmann_nz_lens} have found that uncertainties in the $n(z)$ width are negligible provided that the mean of the distribution is known to sufficient accuracy in a Euclid DR1 3$\times$2pt analysis. In contrast, our results show that dark-energy constraints degrade by an additional 9\% due to uncertainties in the width. The discrepancy is likely caused by differences in the adopted priors for the width parameters. \cite{bertmann_nz_lens} assume $0.1$--$0.8\%$ knowledge of the width, while we adopt a more conservative 1\% uncertainty. Hence, uncertainties in the width may become non-negligible once the lens \nz{} is not calibrated to sub-percent accuracy. 

\section{Reconstruction with ensemble-based priors}\label{subsec:flows}

The biases reported in Appendix \ref{appendix_A} may not be an indictment of the GP
reconstruction itself but of the priors we place on its hyperparameters. The empirical priors of
Section~\ref{subsec:gp_priors} were tuned by visual inspection to be deliberately conservative, and
they consequently grant the multi-node models enough freedom to generate \nz{} realisations that no
realistic calibration pipeline would produce. Those unrealistically large \nz{} uncertainties are
then propagated faithfully, which is what the method is designed to do, into the cosmological and
nuisance posteriors. The fix may therefore not be to reduce the flexibility of the GP but to replace
the hand-tuned prior with one learned from data, which we attempt to achieve using normalising flows (see Section \ref{ensemble_priors} and Fig. \ref{fig:samples_GP}). Here, we present the results from \txtp{} normalising flow chains with three interior training nodes, reducing the number of parameters compared to the four- or five-node configurations. The resulting loss of flexibility is compensated by the per-bin correlation length in the flow parameterisation. The biases and FoMs referred to in this section are reported in Appendix \ref{appendix_A} and \ref{Appendix_B}, respectively.

\subsection{Cosmological constraints}

The left panel of Fig. \ref{fig:flow_posteriors} shows the two-dimensional marginalised posterior distributions of $\Omega_\text{m}$ and $S_8$ in a \LCDM{} cosmology when adopting the flow prior and three training nodes. In \LCDM{}, the flow prior increases FoM($\Omega_\text{m}$, $S_8$) by a factor of 1.4 and 1.2 compared to the empirical prior and the S\&W case, which may be a result of the tighter constraints imposed on \nz{}. Along with this FoM increase, the flow-prior chain introduces a somewhat larger and opposite bias in $\Omega_\text{m}$ with respect to the empirical-prior chain as a consequence of a displaced posterior and a smaller standard deviation. 

\begin{figure}[h!]
    \centering
    \includegraphics[width=\linewidth]{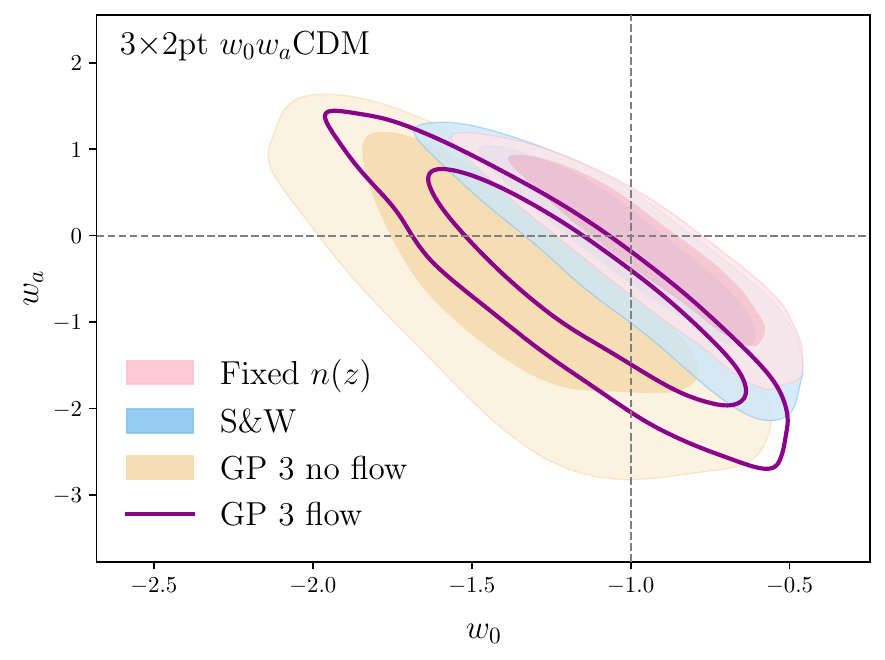}
    \caption{Marginalised two-dimensional ($w_0$-$w_a$) posteriors when adopting ensemble-based (unfilled purple contour) and empirical (filled yellow contour) priors in the three-node configuration. The S\&W and fixed-\nz{} cases are shown for reference and dashed grey lines represent the input values. The posteriors exhibit a pattern similar to that in the right panel of Fig. \ref{fig:flow_posteriors}: the flow contours align more closely with the S\&W case than the empirical-prior contours, although some discrepancy remains.}
    \label{fig:w0wa_flow}
\end{figure}

Similar to our conclusions from the previous section, the differences in \LCDM{} are minor relative to \wowa{}. Compared to the empirical prior, the flow prior moves the \wowa{} ($\Omega_\text{m}$-$S_8$) posterior (right panel of Fig. \ref{fig:flow_posteriors}) somewhat closer to the fiducial, but a significant bias remains. The flow prior increases FoM($\Omega_\text{m}$, $S_8$) by an approximate factor of 1.3 and 1.2 with respect to the empirical prior and the S\&W case, respectively. We observe comparable behaviour in the ($w_0$-$w_a$)-posteriors in Fig. \ref{fig:w0wa_flow}: the flow-chain contours shrink and move towards the S\&W (and fixed-\nz{}) contours and input values, but a discrepancy remains. Thus, although the ensemble-based prior reduces the bias relative to the empirical prior in \wowa{}, the flow posteriors still converge toward a region of parameter space displaced from the fiducial cosmology. The closer agreement across chains in \LCDM{} highlights the greater sensitivity of \wowa{} constraints to variations in \nz{}, consistent with the empirical-prior results discussed in the previous section. 

The Bayesian evidences of the flow chains, $\ln \mathcal{Z} \approx -314.1$ for \LCDM{} and $\ln \mathcal{Z} \approx -309.9$ for \wowa{}, suggest that the ensemble-based GP model is disfavoured by the data relative to the empirical-prior and reference configuration. This may be explained by the per-bin correlation length adopted in the flow configuration, rather than the single global correlation length used in the empirical-prior chains. According to the evidences, this more complex and flexible extension of the \nz{} model is not required by the data, as expected given that the data were generated from a perfect, noiseless \nz{}. The low \LCDM{} evidence is consistent with the total bias accumulated by the flow chain in \LCDM{} ($B = 1.14$), which is the largest overall bias in this cosmology. In contrast, the flow prior reduces the bias from $B = 1.41$ to $B=1.00$ in the three-node GP configuration in \wowa{}, suggesting that the more flexible, data-driven approach is more effective at reducing bias in \wowa{} than in \LCDM{}. We note, however, that the other GP and no-GP models yield lower biases in \wowa{} than the three-node flow configuration, indicating that the latter does not provide the best overall performance. 

\subsection{Origin of remaining discrepancies}
In an attempt to reduce the bias accumulated by the empirical-prior GP models, we implemented the normalising flow ensemble-based prior, limiting the freedom of the on-the-fly GP reconstructions in a data-driven manner. This leaves us with the question as to why the addition of normalising flows still yields substantial biases, which are, in some cases, even larger than the empirical-prior biases. Here, we discuss a few potential reasons for these discrepancies. 

A natural explanation is that the flow chains, despite the more restrictive priors, converge towards a GP hyperparameter region away from the fiducial \nz{}. The right panel of Fig. \ref{fig:preferred_nz_w0wa} shows that this is not the case, since the preferred \wowa{} GP flow reconstruction follows the original \nz{} almost perfectly. The \LCDM{} counterpart exhibits similar behaviour. We infer from this that the ensemble-based prior has successfully restricted the freedom of the GP reconstructions and that offsets from the fiducial \nz{} are not the origin of the observed discrepancies. 

Another possible cause is that the flow prior interacts differently with cosmological and nuisance parameter degeneracies. For instance, a degeneracy between cosmological or nuisance parameters and the \nz{} reconstructions may be broken by the tighter prior placed on the GP hyperparameters. As a consequence, shifts that can be compensated by changes in the \nz{} reconstructions allowed by the empirical prior may no longer be compensated when adopting the more rigid ensemble-based prior. This would lead to larger biases in cosmological or nuisance parameters, and we observe both for the \LCDM{} flow configuration (top panel of Fig. \ref{fig:biases}).  Interestingly, the \LCDM{} flow chain follows a distinct, biased trajectory in the preferred nuisance parameter space, suggesting that cosmological shifts are increasingly compensated by nuisance parameters rather than GP hyperparameters. In contrast, the \wowa{} bias trajectory of the flow chain across the parameter space is comparable to the other GP chains, indicating that the additional dark-energy degrees of freedom have altered the degeneracy structure with respect to \LCDM{}. 

Although their precise origin remains uncertain, the discrepancies between the ensemble-based prior constraints and the fiducial parameter values underline the complexity of propagating \nz{} uncertainties. This complexity reflects the interplay between the adopted \nz{} uncertainty model, the assumed cosmology, and the parameter degeneracy structure that both introduce. Consequently, even \nz{} reconstructions that closely reproduce the input distributions may yield biased cosmological constraints, particularly in the \wowa{} cosmology. These findings highlight the importance of carefully selecting \nz{} uncertainty propagation methods for upcoming Stage-IV dark energy surveys, where unbiased evolving dark energy constraints are a main objective. 

\section{Conclusions}\label{sec:conclusions}

As photometric surveys move from Stage III to Stage IV, the statistical uncertainties on the
\txtp{} observables shrink to the point where the treatment of systematics, and of photometric
redshift calibration in particular, sets the accuracy of the inferred cosmology. In this work we
have introduced a data-driven method for propagating redshift-distribution uncertainties as nuisance parameters in the likelihood analysis that makes
no assumption about the functional form of \nz{}: the distribution in each tomographic bin is
reconstructed on the fly with a Gaussian process (GP), conditioned on a small number of training nodes
whose positions and amplitudes are sampled jointly with the cosmological and nuisance parameters.
Positivity is guaranteed by regressing on $\sqrt{n(z)}$, the effective resolution follows the
narrowing of the low-redshift bins through a logarithmic coordinate, and the reconstruction is
inexpensive enough to be evaluated at every likelihood call. We impose and compare two different sets of priors on the GP hyperparameters: empirically informed priors and ensemble-based normalising flow priors. The empirical model uses a single global correlation length, while the flow model adopts bin-specific correlation lengths. We apply both methods to synthetic \textit{Euclid} DR1-like \txtp{} data vectors in a Bayesian inference framework. The GP implementation of the \nz{} is released in the python package
\texttt{binz}, and is built to work smoothly with the public \texttt{cloe-org} ecosystem.

Our central result is
that \nz{} uncertainty modelling is inherently cosmology dependent. When adopting empirical priors in \LCDM{}, the multi-node GP
reconstructions agree with each other and with the shifts-and-widths (S\&W) and fixed-\nz{} baselines within uncertainty, the preferred
reconstructions track the input distribution, and the total parameter bias decreases monotonically
as training nodes are added. Freeing $w_0$ and $w_a$ changes this picture qualitatively rather than
quantitatively: the three- and five-node configurations move to a distinctly different region of
parameter space, the total biases grow for every \nz{} model, and the optimal GP configuration is
no longer the most flexible one but the four-node case, which nevertheless prefers a source \nz{}
that departs visibly from the input, with the discrepancy absorbed by the per-bin magnification
biases rather than by the cosmology. The \nz{} shape favoured by the data is therefore itself a
function of the cosmological model assumed, which is a direct problem for any method whose basis is derived under a fixed cosmology (e.g., mode projection, PCA marginalisation).

When limiting the freedom of the GPs by imposing ensemble-based flow priors, the preferred reconstructions follow the input \nz{} closely in both \LCDM{} and \wowa{}. Despite these similar reconstructions, the cosmological constraints differ significantly between the two cosmologies, similarly to the empirical prior results. In \LCDM{}, the flow prior tightens the ($\Omega_\text{m}$-$S_8$) posterior, but the recovered parameter values remain consistent across \nz{} uncertainty models. When including $w_0$ and $w_a$ on the other hand, although the flow prior reduces the bias with respect to the empirical prior, the posteriors are shifted with respect to the fiducial. The remaining discrepancies between the fiducial parameter values and the flow constraints highlight the complex degeneracies between the \nz{}, cosmological and nuisance parameters. Moreover, cosmological constraints obtained in a \wowa{} cosmology appear more sensitive to \nz{} variations than those acquired in \LCDM{}.

Another methodological conclusion is that constraining power alone is a poor criterion to compare \nz{} models. Several GP configurations yield tighter posteriors than the shifts-and-widths baseline while being more biased, and the flexible models are consistently disfavoured in Bayesian evidence, which correctly penalises the large prior volume they open. Bias
and evidence must be reported alongside any figure of merit.

It is worth mentioning that we do not claim that the GP models in their present form are ready for a cosmological analysis. Although the ensemble-based prior successfully reduces the GP freedom, our \nz{} uncertainty models yield substantial biases. Even in their current state, however, the GP reconstructions have served their diagnostic purpose: 
Their flexibility exposes a sensitivity of dark-energy constraints to redshift-distribution
modelling that rigid parameterisations cannot reveal, precisely because those parameterisations
never propose the deformations that trigger it. Yet, our recommendation for Stage-IV analyses follows
directly: \nz{} uncertainty models should be validated separately in every cosmological model in
which they will be used, and the validation should be carried out in \wowa{} before any
dark-energy constraint derived with them is quoted. 

Several directions follow naturally from this work. The first is to add external
information that can self-calibrate the \nz{}. Cross-correlations with CMB lensing are a
natural candidate, since they constrain the redshift distributions independently of the
\txtp{} data and could break the degeneracies between \nz{}, cosmological and nuisance
parameters that we identified in \wowa{}. The second is to benchmark the GP reconstruction
against other recent approaches to redshift-distribution uncertainty, in preparation for the
first cosmological data releases of Stage-IV photometric surveys. Finally, as the dimension
of the sampled parameter space grows with the flexibility of the \nz{} model, we plan to develop summary statistics for bias and constraining power that remain meaningful in high-dimensional spaces. These would complement the figures of merit and Bayesian evidence used here.

\section*{Acknowledgments}
We acknowledge the use of software developed and maintained by the \texttt{cloe-org} team, an open-source ecosystem for cosmological modelling and statistical inference for large-scale structure adopted by the Euclid Consortium. This work used \texttt{cloelib} and \texttt{cloelike}, which are publicly available at
\href{https://github.com/cloe-org}{https://github.com/cloe-org}. GCH acknowledges that this project is part of the project UNICORN with file number VI.Veni.242.110 of the research programme Talent Programme Veni Science domain 2024 which is (partly) financed by the Dutch Research Council (NWO) under the grant \url{https://doi.org/10.61686/ZCPQI32997}. HH acknowledges support from the European Research Council (ERC) under the
European Union’s Horizon 2020 research and innovation program (Grant Agreement No. 10105399). We acknowledge useful discussions with W. Hartley, N. Tessore, Y. H. Zhang. We acknowledge the EuroHPC Joint Undertaking for awarding this project access to the EuroHPC supercomputer LEONARDO, hosted by CINECA (Italy) and the LEONARDO consortium, through a EuroHPC Extreme Access call.

\bibliographystyle{aasjournal}

\bibliography{oja_template}

\setcounter{figure}{0}
\renewcommand{\thefigure}{A\arabic{figure}}
\begin{figure*}[t]
    \centering
    \includegraphics[width=0.92\textwidth]{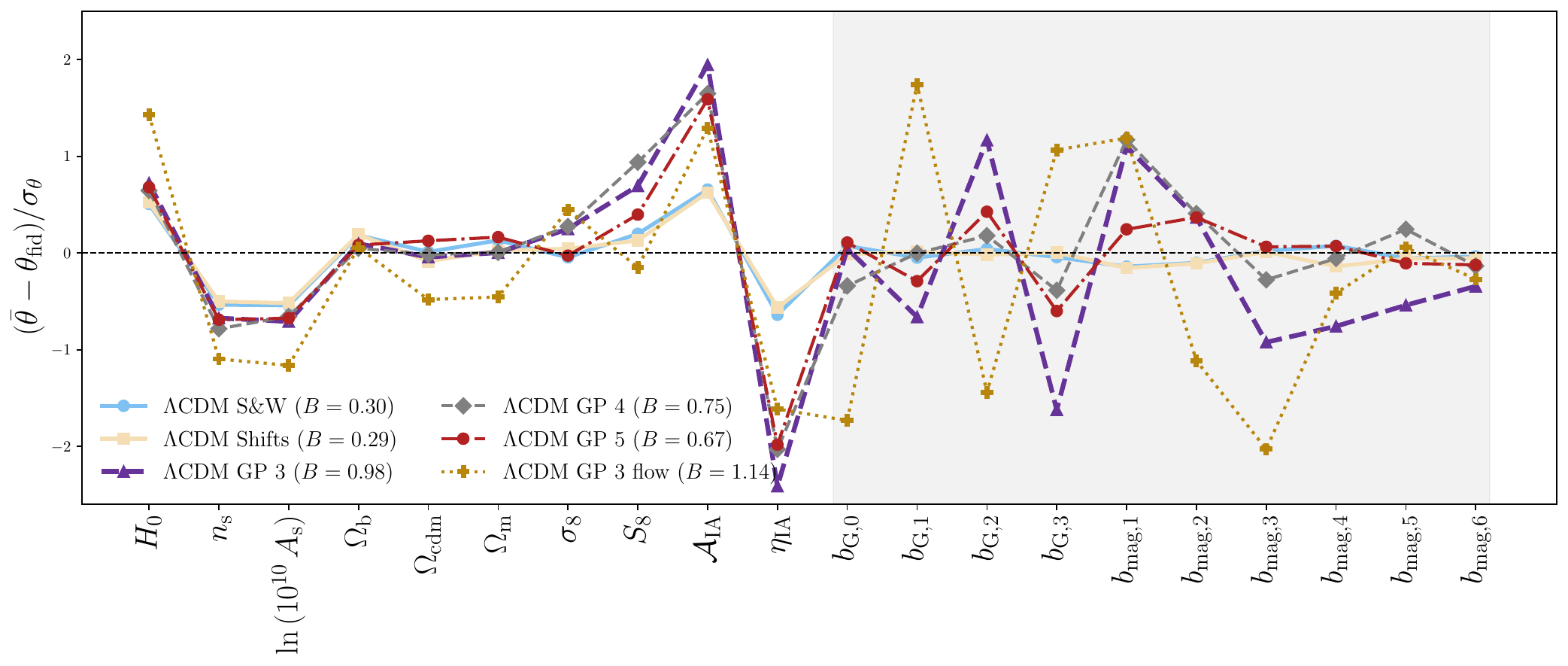}
    \\[0.3cm]
    \includegraphics[width=0.92\textwidth]{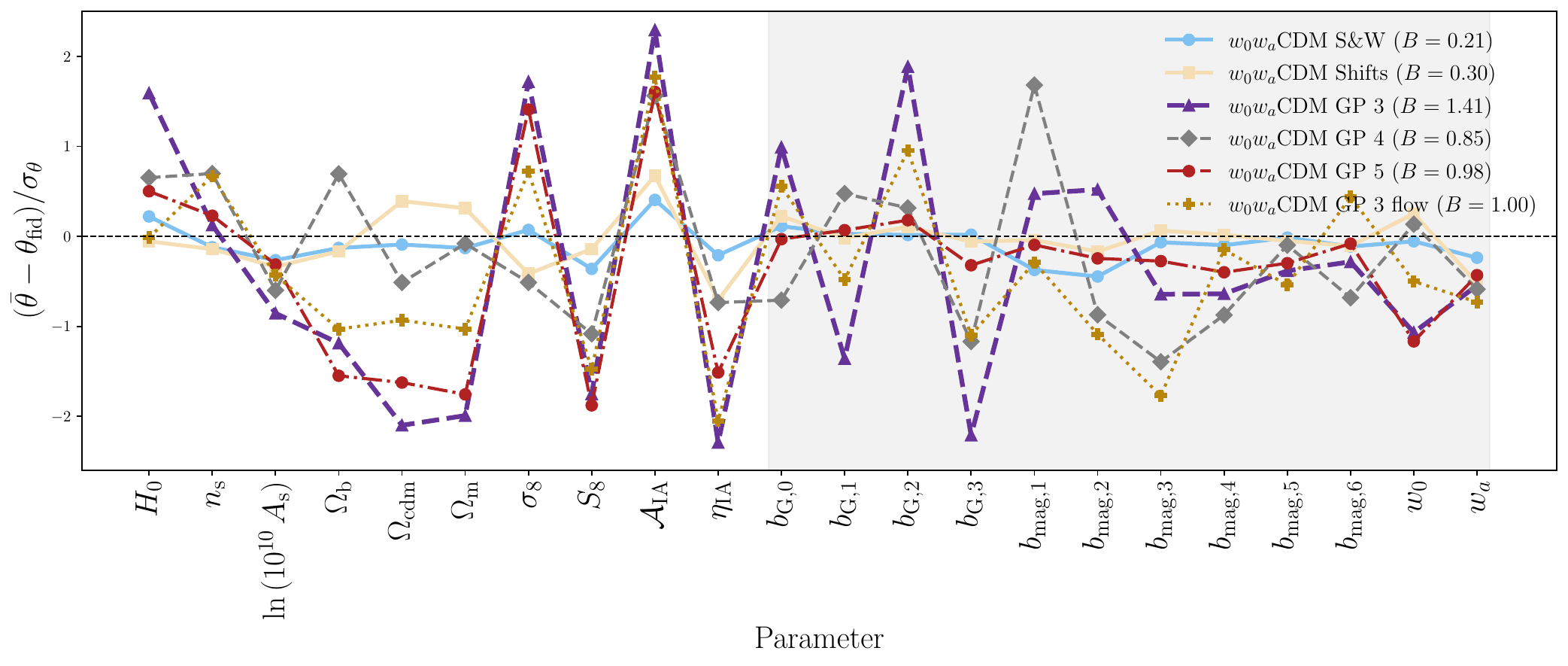}
    \caption{Bias of the cosmological and nuisance parameters, Eq.~\eqref{eq:bias}, for the
    different \nz{} uncertainty models in a \LCDM{} (top) and \wowa{} (bottom) \txtp{} analysis.
    The total bias $B = \big(\frac{1}{N}\sum_i b_i^2\big)^{1/2}$ of each chain is given in the legend. The per-bin
    multiplicative shear biases are omitted, as their tight Gaussian prior makes their bias
    uninformative. The shaded regions indicate redshift-dependent parameters, which show larger fluctuations in bias across \nz{} uncertainty models than the unshaded regions in \LCDM{}, whereas fluctuations in \wowa{} are relatively large across the entire parameter space. The bias ordering of the GP configurations is monotonic in the number of nodes
    in \LCDM{} but not in \wowa{}.}
    \label{fig:biases}
\end{figure*}

\appendix{}

\section{Parameter biases} \label{appendix_A}

Figure~\ref{fig:biases} collects the per-parameter biases of Eq.~\eqref{eq:bias} for the
shifts-only, S\&W and three multi-node GP models, in \LCDM{} (top) and \wowa{} (bottom), with the
total bias $B$ of each chain quoted in the legend. The shaded region represents redshift-dependent parameters. Three features stand out.

First, the ordering of the models changes between cosmologies. In \LCDM{} the total bias decreases
monotonically with the number of training nodes, consistent with the fact that a more flexible GP
recovers the input \nz{} from which the data were generated more faithfully. In contrast, in \wowa{} the four-node configuration attains the smallest GP bias, while the three-
and five-node models incur the largest, particularly in $\Omega_{\rm cdm}$, $\sigma_8$ and $S_8$.
The configuration that is optimal in \LCDM{} is therefore not the one that is optimal in \wowa{}.

Second, the overall bias level is higher in \wowa{} for every \nz{} model, reflecting the extra
degeneracy directions opened by $w_0$ and $w_a$, along which a mis-modelled \nz{} can propagate.
The three- and five-node configurations show a large bias in $w_0$ specifically, whereas the bias
in $w_a$ is comparatively stable across \nz{} models. 

Third, in both cosmologies the largest differences between GP and non-GP runs appear in the
nuisance sector rather than in the cosmological one. Specifically, differences  arise most prominently in the IA parameters in \LCDM{} and in the per-bin magnification biases for the four node model in \wowa{}. The IA
parameters are thus correlated with the \nz{} uncertainty model. Finally, the bias in the redshift-dependent parameters (grey area) varies substantially across \nz{} uncertainty methods in both cosmologies, whereas for parameters that are largely independent of redshift, the variation is more pronounced in \wowa{}. The choice of \nz{} uncertainty method in \wowa{} thus affects not only redshift-dependent parameters but also those that are largely independent of redshift.

\setcounter{figure}{0}
\renewcommand{\thefigure}{B\arabic{figure}}

\begin{figure*}[t]
    \centering
    \includegraphics[width=0.92\textwidth]{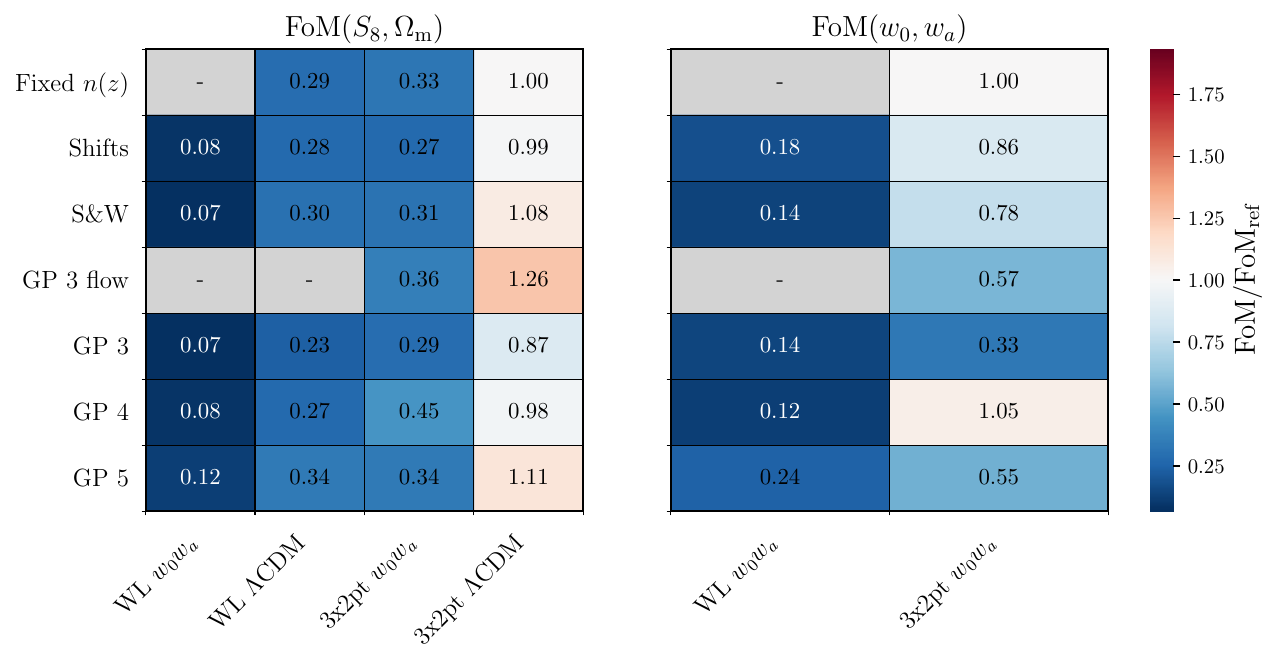}
    \caption{Figures of merit of the different \nz{} uncertainty models relative to the fixed-\nz{}
    \txtp{} reference, for $(\Omega_{\rm m}, S_8)$ (left) and $(w_0, w_a)$ (right), across
    cosmologies and probe combinations. Redder cells indicate more constraining power than the
    reference, bluer cells less, and grey cells represent configurations in which no chain was run. The apparent gain of the more flexible GP models in
    $(\Omega_{\rm m}, S_8)$ must be read alongside Fig.~\ref{fig:biases}: a narrower posterior
    centred away from the input is not an improvement.}
    \label{fig:fom}
\end{figure*}
\section{Parameter constraints} \label{Appendix_B}
The constraining power tells a complementary story to the biases. Figure~\ref{fig:fom} shows the FoM of each
model relative to the fixed-\nz{} \txtp{} reference, for $(\Omega_{\rm m}, S_8)$ (left) and $(w_0, w_a)$
(right), across cosmologies and probe combinations. For $(\Omega_{\rm m}, S_8)$ the dominant trend
is horizontal: the \txtp{} \LCDM{} configuration is by far the most constraining, WL-only \wowa{}
the least, and \txtp{} \wowa{} and WL-only \LCDM{} are comparable. Opening the dark-energy sector thus
costs roughly as much constraining power as discarding galaxy clustering and galaxy--galaxy lensing
altogether. Within \txtp{} \LCDM{}, the FoM increases with the number of training nodes,
with the five-node model reaching a factor $\approx 1.1$ higher than the fixed-\nz{} reference. Taken together
with the top panel of Fig.~\ref{fig:biases}, this increase does not indicate a more accurate
measurement, since the GP runs are simultaneously more biased in $S_8$. Constraining power without an accompanying bias diagnostic is therefore a misleading
figure of merit for \nz{} model comparison.

For the dark-energy FoM the four-node GP model is the most constraining \txtp{} configuration,
followed by a fixed \nz{} and shifts-only model, with the three- and five-node models the weakest, which mirrors
Fig.~\ref{fig:biases}. Here the four-node bias in $w_0$ and $w_a$ is comparable to the
non-GP runs, so the large FoM is not simply the signature of a displaced posterior; it follows
instead from the narrow region of GP hyperparameter space that the data select in this
configuration, whose small \nz{} uncertainty propagates directly into tight dark-energy
constraints. Finally, the shifts-only model attains a higher dark-energy FoM than S\&W, which
quantifies how much of the $w_0$--$w_a$ constraining power is lost to width uncertainties alone:
FoM$(w_0, w_a)$ drops from $\approx 11.8$ to $\approx 10.7$, a $\sim 9\%$ reduction, when the widths are
allowed to vary in addition to the means.

\setcounter{figure}{0}
\renewcommand{\thefigure}{C\arabic{figure}}
\section{Preferred GP reconstructions} \label{Appendix_C}
\begin{figure*}[t]
    \centering

    \begin{subfigure}{0.46\textwidth}
        \centering
        \includegraphics[width=\linewidth]{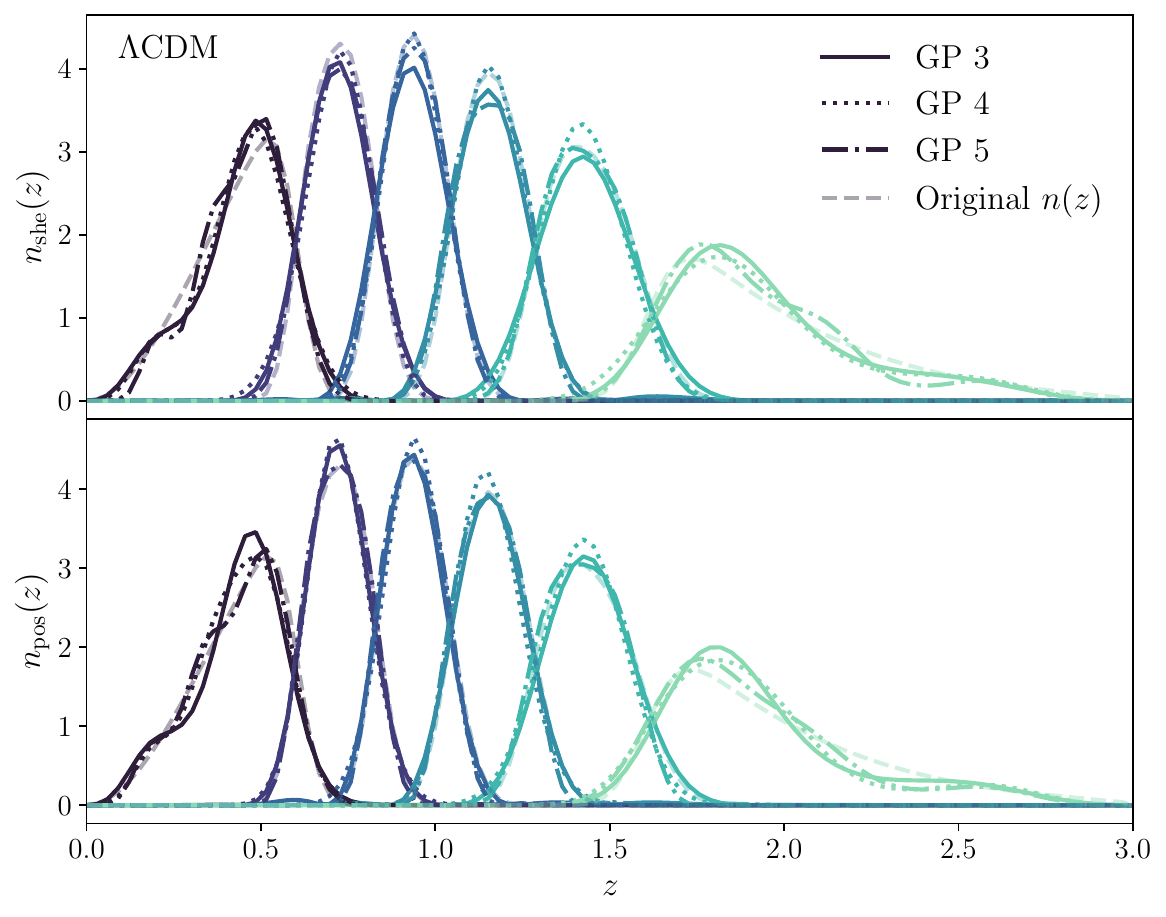}
    \end{subfigure}
    \begin{subfigure}{0.46\textwidth}
        \centering
        \includegraphics[width=\linewidth]{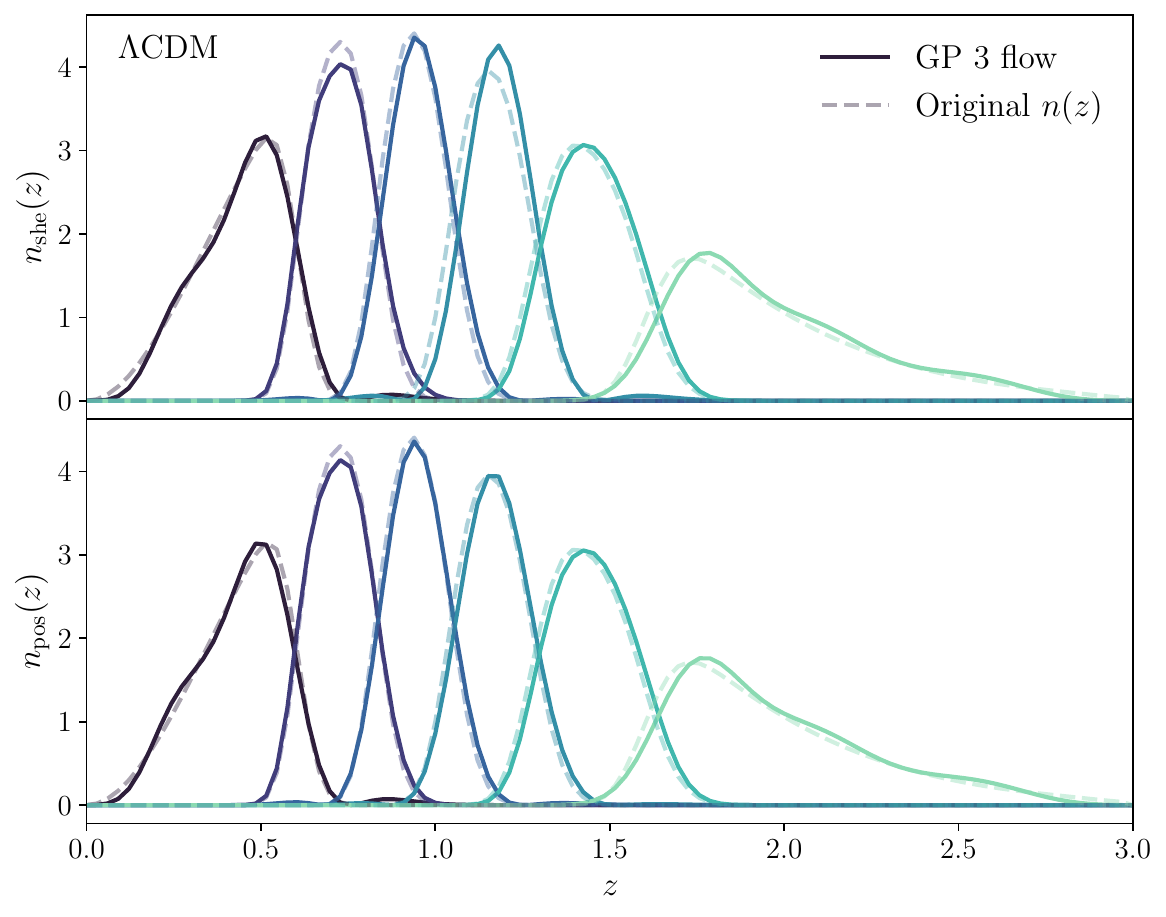}
    \end{subfigure}

    \begin{subfigure}{0.46\textwidth}
        \centering
        \includegraphics[width=\linewidth]{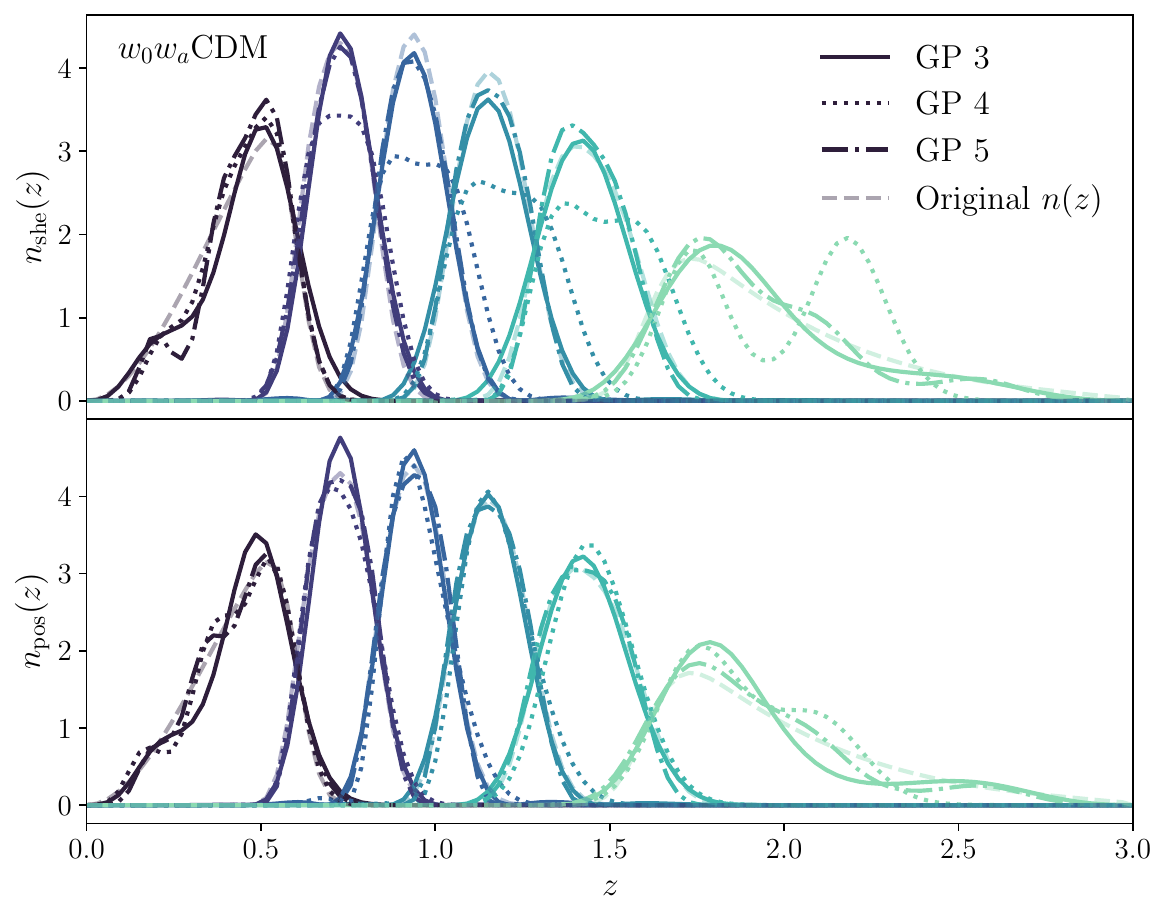}
    \end{subfigure}
    \begin{subfigure}{0.46\textwidth}
        \centering
        \includegraphics[width=\linewidth]{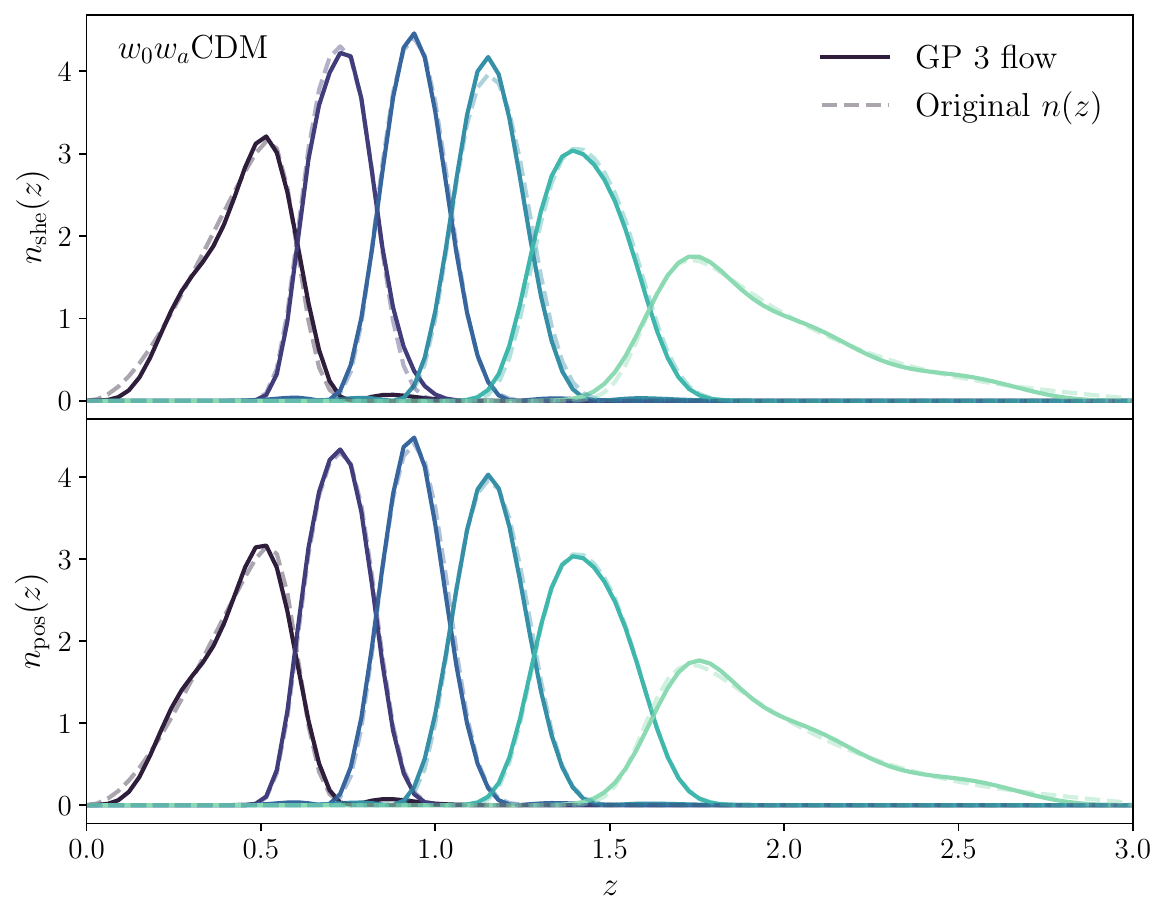}
    \end{subfigure}
    
    \caption{Normalised source and lens \nz{} reconstructions preferred by the data
    in the \txtp{} \LCDM{} (top) and \wowa{} (bottom) analyses, evaluated at the marginalised posterior means of the GP
    hyperparameters with empirical (left) and ensemble-based (right) priors, for three (solid), four (dotted) and five (dot-dashed) training nodes. The
    input \nz{} is shown as a dashed line. The empirical four-node \wowa{} source reconstruction is bimodal in the
    highest-redshift bin and systematically lower and broader in the central bins. The flow-based reconstructions follow the original \nz{} closely, more so than their empirical-prior counterparts. 
    }
    \label{fig:preferred_nz_w0wa}
\end{figure*}

Figure \ref{fig:preferred_nz_w0wa} shows the source and
lens reconstructions preferred by the data in a \wowa{} cosmology, built from the marginalised posterior means of the GP
hyperparameters. The three- and five-node reconstructions track the input distribution closely, and
all lens reconstructions are broadly consistent with it apart from the highest-redshift bin. The
four-node \emph{source} reconstruction, however, departs markedly from the input. Namely, the central bins
are systematically lower and broader, and the distribution in bin 6 becomes bimodal, suggesting that the
data preferred to split the source sample into seven bins rather than six. That such a distorted
reconstruction still yields near-reference cosmological constraints is explained by shifts in the per-bin
magnification biases that compensate for the $n(z)$ deviations, rather than by shifts in the cosmological parameters. This is precisely the
kind of degeneracy that a low-dimensional \nz{} parameterisation cannot expose, because it never
proposes such a deformation in the first place.

\begin{figure}[h!]
    \centering
    \includegraphics[width=\linewidth]{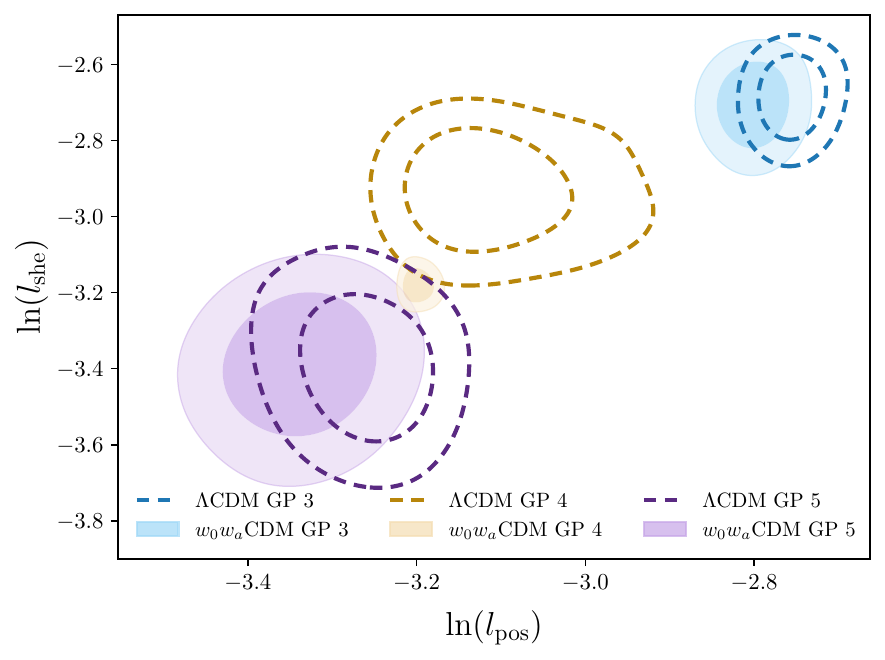}
    \caption{Marginalised two-dimensional posteriors of the source and lens correlation lengths,
    $\ln l_{\rm she}$ and $\ln l_{\rm pos}$, from \txtp{} \LCDM{} (dashed, unfilled) and \wowa{}
    (filled) GP chains with three, four and five training nodes. The correlation length decreases
    with node density, as expected; the four-node \wowa{} posterior is dramatically tighter than
    its \LCDM{} counterpart.}
    \label{fig:hyper}
\end{figure}

Figure~\ref{fig:hyper} shows the joint posterior of the source and lens correlation lengths,
$\ln l_{\rm she}$ and $\ln l_{\rm pos}$, for the three GP configurations in both cosmologies. As
expected, the preferred correlation length decreases as nodes are added and their separation
shrinks. The noticeable feature is the four-node case. Relative to its \LCDM{} counterpart, the
\wowa{} posterior is tighter by a factor of $5.7$ in $l_{\rm pos}$ and $4.0$ in $l_{\rm she}$, with
the mean shifted by $-1.3\sigma$ and $-2.4\sigma$ respectively. In other words, the combination of
four nodes with the additional dark-energy degrees of freedom produces a parameterisation so
well matched to the data that only a narrow range of smoothness scales fits, and the resulting
contraction of the \nz{} sector propagates into unusually tight---and, as
Fig.~\ref{fig:preferred_nz_w0wa} shows, incorrectly centred---nuisance constraints. We find no
significant degeneracies between GP hyperparameters and nuisance parameters, which rules out
degeneracy breaking as the origin of the extra constraining power.

\end{document}